\documentclass[fleqn, usenatbib]{mnras}

\usepackage[T1]{fontenc}
\usepackage[utf8]{inputenc}

\usepackage{amsmath} 
\usepackage{amsfonts} 
\usepackage{amsbsy} 
\usepackage{amssymb} 
\usepackage{mathtools} 
\usepackage{stmaryrd} 
\usepackage{stackengine}  

\usepackage{graphicx} 
\usepackage{subcaption} 

\usepackage{acronym} 
\usepackage{soul} 
\usepackage{xcolor} 

\makeatletter
\newcommand\footnoteref[1]{\protected@xdef\@thefnmark{\ref{#1}}\@footnotemark}
\makeatother

\newacro{PPD}{proto-planetary disc}
\newcommand{\PPD}{\ac{PPD}}
\newacroplural{PPD}{proto-planetary discs}
\newcommand{\PPDs}{\acp{PPD}}

\newacro{SI}{streaming instability}
\newcommand{\SI}{\ac{SI}}
\newacroplural{SI}{streaming instabilities}
\newcommand{\SIs}{\acp{SI}}

\newacro{RWI}{Rossby wave instability}
\newcommand{\RWI}{\ac{RWI}}

\newacro{SBI}{sub-critical baroclininc instability}

\newacro{COS}{convective over-stability}

\newacro{RDI}{resonant drag instability}
\newcommand{\RDI}{\ac{RDI}}
\newacroplural{RDI}{resonant drag instabilities}
\newcommand{\RDIs}{\acp{RDI}}

\newacro{EI}{elliptical instability}
\newcommand{\EI}{\ac{EI}}

\newacro{ODE}{ordinary differential equation}
\newcommand{\ODE}{\ac{ODE}}
\newacroplural{ODE}{ordinary differential equations}
\newcommand{\ODEs}{\acp{ODE}}

\newacro{PDE}{partial differential equation}

\newacroplural{PDE}{partial differential equations}
\newcommand{\PDEs}{\acp{PDE}}

\newacro{IVP}{initial value problem}
\newcommand{\IVP}{\ac{IVP}}

\newacro{DAE}{differential-algebraic equation}
\newcommand{\DAE}{\ac{DAE}}

\newcommand{\nameofmycode}{LAVA} 

\newacro{TVA}{terminal velocity approximation}

\newacro{MMSN}{minimum mass solar nebulae}

\newacro{ALMA}{Atacama Large Millimeter Array}

\newacro{VLT}{Very Large Telescope}

\newcommand{\bb}[1]{\mbox{\boldmath{$#1$}}}

\newcommand{\bnabla}{\bb{\nabla}} 
\newcommand{\rd}{\mathrm{d}} 
\newcommand{\Dt}{\mathrm{D}_{t}} 

\newcommand{\bcdot}{\, \mbox{\boldmath{$\cdot$}} \,}

\newcommand{\ex}{\ensuremath{\mathbf{e}_{x}}} 
\newcommand{\ey}{\ensuremath{\mathbf{e}_{y}}} 
\newcommand{\ez}{\ensuremath{\mathbf{e}_{z}}} 

\newcommand{\bx}{\mathbf{x}} 
\newcommand{\bu}{\mathbf{u}} 
\newcommand{\bk}{\mathbf{k}} 

\newcommand{\e}{\mathrm{e}} 

\newcommand{\dtg}{\mu} 
\newcommand{\St}{\text{St}} 

\title[The effect of dust on vortices II]{The effect of dust on vortices II: Streaming instabilities}
\author[N.~Magnan \& H.~Latter]{
    Nathan Magnan$^{1, 2}$
    and Henrik N.~Latter$^{2}$\thanks{Contact e-mail: \href{mailto:hl278@cam.ac.uk}{hl278@cam.ac.uk}} 
    \\
    $^{1}$ Laboratoire Lagrange, Observatoire de la Côte d'Azur, Université de la Côte d'Azur, Nice, France\\
    $^{2}$ DAMTP, University of Cambridge, Cambridge, UK
    }

\begin{document}

\label{firstpage}
\pagerange{\pageref{firstpage}--\pageref{lastpage}}

\maketitle

\begin{abstract}
    One of the main questions in planet formation theory is how to cross the metre-scale barrier. In this two-part series, we assess the merits of vortex-based theories by investigating the effect of backreacting dust on vortices. Specifically, this second paper focuses on the ‘turbulent’ vortex theory, according to which the streaming instability~(SI) might be active in vortices. We re-purpose the dusty vortex models derived in paper~I as background flows for a linear stability analysis. To simplify the perturbation equations, we build an analogue of the shearing box that follows vortex streamlines instead of Keplerian orbits. This allows us to study the evolution of small wavelength perturbations. We find that inertial waves and dust density waves can propagate in vortices, but that they are not sinusoidal in time. We then extend resonant drag instability theory to these non-modal waves. This allows us to demonstrate that a close cousin of the SI remains active in vortices, a result that greatly strengthens the case for vortex-induced planetesimal formation. Our results also complement past simulations -- which showed that the dust's backreaction makes vortices unstable -- by providing insights into the nature of (some of) the unstable modes. The caveat is that our work is restricted to the limit of dilute well-coupled dust and to the linear phase of the instability. Finally, our `vortex SI’ extends to 2D. We explain the mechanism of this `zonal flow RDI', but remain unsure whether it is the unknown instability seen in 2D vortex simulations.
\end{abstract}


\section{Introduction}
\label{sec:Introduction}

\phantom{\PPD}
\vspace{-1 \baselineskip}

\noindent ALMA detected large-scale vortices in several protoplanetary discs (\PPDs\ -- \citealt{Bae+2023}). These vortices interest planet formation theorists because they tend to capture large amounts of pebbles \citep{BargeSommeria1995, AdamsWatkins1995, Tanga+1996, Chavanis2000}, and if the pebble density becomes larger than the Hill density, the vortex core can collapse and form planetesimals. This would provide a bridge over the metre-scale barrier.

However, we argued in paper I that laminar vortices are unlikely to concentrate dust sufficiently to trigger gravitational collapse. Indeed, dust affects the shape of vortices, bringing them to a region of parameter space where the \EI\ is active \citep{LesurPapaloizou2009}. This instability destroys the vortex before laminar concentration can bring the dust to the Hill density. Therefore, an additional and faster dust concentration mechanism is required.

\phantom{\SI}
\vspace{-1 \baselineskip}

The streaming instability (\SI\ -- \citealt{YoudinGoodman2005}) is a promising way to create strong dust overdensities. Unfortunately, it requires a high metallicity \citep{Carrera+2015, Yang+2017, LiYoudin2021}, similar-sized particles \citep{Krapp+2019, YangZhu2021, Paardekooper+2020, PaardekooperAly2025}, little to no turbulent diffusion \citep{ChenLin2020, Umurhan+2020, Gole+2020, Lim+2024}, and a weak radial pressure gradient \citep{BaiStone2010, SekiyaOnishi2018, Baronett+2024}.

Anticyclonic vortices selectively trap particles of a certain size, so given enough time they can provide both a high density of pebbles and a narrow distribution in particle size. Furthermore, they are thought to be isolated from the disc's turbulence \citep{KlahrBodenheimer2006}, and weak anticyclones are pressure maxima, so the pressure gradient can be as small as required. All of this advocates for a `turbulent' vortex pathway to planet formation.

Unfortunately, fluid instabilities are notoriously fragile to modifications of their background flow (see, \textit{e.g.}, \citealt{Kelly1992}), the \SI\ is a particularly finicky instability \citep{Magnan2024b}, and vortex flows differ significantly from Keplerian flows. Therefore, the instability may not be active in vortices. The goal of the present paper is to adress this concern.

Several teams have already tried to address this problem using numerical experiments \citep{Fu+2014, CrnkovicRubsamen+2015, Raettig+2015, Surville+2016, Miranda+2017, Lyra+2018, SurvilleMayer2019, Raettig+2021, Lovascio+2022}. Their setups vary (2D or 3D vortices, \textit{ab initio} or sustained vortices, \textit{etc.}), but apart from \cite{Miranda+2017}, they all agree that the dust's backreaction triggers an instability. It cannot be the heavy-core instability of \cite{ChangOishi2010} nor the parametric instability of \cite{RailtonPapaloizou2014} because both require perfectly entrained dust.

Unfortunately, the conditions required to trigger this instability remain unclear. According to \cite{Fu+2014} and \cite{Raettig+2015}, it appears as soon as the dust-to-gas ratio reaches unity in the vortex’s core. \cite{Lovascio+2022} adds that the dust also needs to drift relative to the gas. All of this may be in disagreement with \cite{Surville+2016}, whose instability first appears in damped vortex cores (where the dust-to-gas drift is minimal) before spreading outwards (to regions where there is less dust). Finally, \cite{Lovascio+2022} find that vortices with highly subsonic velocity perturbations relative to the Keplerian background are stable.

The instability seems to appear at large scales in~\cite{Surville+2016} and~\cite{Fu+2014}. However, \cite{SurvilleMayer2019} report that the vortex-wise azimuthal wavenumber of the fastest-growing mode increases with resolution.

In 2D, the instability eventually destroys the vortex. However, this process takes thousands of orbits in \cite{Fu+2014}, a few hundred orbits in \cite{Surville+2016}, and a dozen orbits in \cite{Raettig+2015}. All three studies assume similar Stoke numbers and similar dust-to-gas ratios, but their vortices differ. This may explain the discrepancy.

But beyond those quantitative mismatches, the main issue with simulations is that they reveal very little about the nature of the instability. This is problematic, because if the instability is a form of \SI, then it probably saturates by forming gravitationally bound clumps. But a strong argument against this is that the instability appears in 2D simulations, whereas the classical \SI\ requires a vertical dimension. And if the instability has a different nature, then it may not form clumps.

Of course, one could adress this question by simulating the instability all the way to saturation, and checking if dust clumps appear. Unfortunately, it seems that the resolutions of all the aforementionned simulations are insufficient. \cite{Fu+2014} show that they have just enough resolution to resolve the laminar dust capture phase, \cite{Raettig+2015} recognise that they cannot resolve all the unstable wavelengths and therefore cannot predict the growth rate of the instability, the maximum dust density is multiplied by ten between \cite{Surville+2016}’s low-resolution and high-resolution runs, \textit{etc.}

The last issue with simulations is their cost. The system is governed by at least four not-necessarily-scalar parameters (the dust-to-gas ratio, the particles' size distribution, the vorticity profile, and the disc's shear rate), so exploring parameter space would be prohibitively expensive.

\phantom{\RDI}
\vspace{-1 \baselineskip}

To overcome some of those issues, we study the linear stability of dust-laden vortices in protoplanetary discs analytically. We start by presenting in \S\ref{sec:Governing_equations} the physical system and its governing equations, and by recalling in \S\ref{sec:Background_flow} the dusty vortex models from paper I. They appear steady on the orbital and vortex turnover timescales, so we can use them as background flows for a linear perturbation analysis in \S\ref{sec:Linear_perturbation_equations}. We then move on the `results' section of the paper. We know that the \SI\ is a resonant drag instability (\RDI\ -- \citealt{SquireHopkins2018b}), so it arises when a gas wave and a dust density wave share the same frequency. This motivates us to study in \S\ref{sec:Waves} the waves that propagate in our vortices. We then leverage those findings in \S\ref{sec:Instability} to show that the dust's backreaction makes vortices unstable, and that the instability is a form of \SI. This represents a major step towards validating the `turbulent' vortex pathway to planet formation. Finally, we explore parameter space in \S\ref{sec:Exploration_of_parameter_space}, we discuss our hypotheses and findings in \S\ref{sec:Discussion}, and we conclude in \S\ref{sec:Conclusion}.

\section{Governing equations}
\label{sec:Governing_equations}

\newcommand{\rhog}{\rho_{g}} 
\newcommand{\bug}{\bu_{g}} 
\newcommand{\rhod}{\rho_{d}} 
\newcommand{\bud}{\bu_{d}} 

We consider a gas and dust mixture, which we model as a two-fluid system. The gas is assumed to be incompressible, and is described by its density~$\rhog$, its velocity~$\bug$, and its pressure~$P$. The dust is assumed to be pressure-less, and is described by its density~$\rhod$ and velocity~$\bud$. The two fluids are coupled by a linear drag force from the gas onto the dust, and its back-reaction from the dust onto the gas.

\newcommand{\eX}{\ensuremath{\mathbf{e}_{X}}} 
\newcommand{\eY}{\ensuremath{\mathbf{e}_{Y}}} 
\newcommand{\eZ}{\ensuremath{\mathbf{e}_{Z}}} 

To account for the \PPD\ context, we work in the shearing box \citep{GoldreichLyndenBell1965, Hawley1995, LatterPapaloizou2017}. The centre of the box orbits at frequency $\Omega$ and radius $r_{0}$. The local Cartesian coordinate system has its $X$-axis oriented in the radial direction, its $Y$-axis in the azimuthal direction, and its $Z$-axis in the direction normal to the disc's plane. The Keplerian flow, the tidal force, and the Coriolis force take their standard forms,
\begin{subequations}
    \label{eq:flow_and_forces_in_shearing_box}
    \begin{align}
        \bu_{\mathrm{sb}} &= - S X \eY , \label{eq:flow_in_shearing_box} \\
        - \bnabla \Phi_{t} &= 2 \Omega S X \eX , \label{eq:forces_in_shearing_box_tidal_potential} \\
        \mathbf{f_{Co}^{\text{s}}} (\bu) &= - 2 \Omega \, \eZ \wedge \bu , \label{eq:forces_in_shearing_box_Coriolis}
    \end{align}
\end{subequations}
where the subscript $\mathrm{sb}$ stands for \textit{shearing box}, ${ S \approx (3/2) \, \Omega }$ is the local shearing rate of the disc, and $\bu$ could represent either the gas or the dust's velocity. 

Note that we ignore the disc's large-scale pressure gradient and the vertical component of gravity. Indeed, our focus is on the midplane layer of the vortex's core.

\newcommand{\bFi}{\mathbf{F_{u}}} 
\newcommand{\bFii}{\mathbf{F_{x}}} 

Note also that the Coriolis force depends on velocity not position, whereas the tidal force depends on position not velocity. We shall group forces of the first kind in a term labelled ${ \bFi }$, and forces of the second kind in ${ \bFii }$. This distinction is useful because only the first group affects linear stability.

Under those assumptions and notations, the mass and momentum conservation equations for gas and dust are
\begin{subequations}
    \label{eq:Navier_Stockes_equations}
    \begin{align}
        \bnabla \bcdot  \bug &= \,\, 0 , \phantom{\frac{1}{1}} \label{eq:continuity_gas} \\
        \partial_{t} \rhod + \bud \, \bcdot \bnabla \rhod  &= - \rhod \, \bnabla \bcdot \bud , \phantom{\frac{1}{1}} \label{eq:continuity_dust} \\
        \partial_{t} \bug \! + \! \bug \bcdot \bnabla \bug \! &= \! - \frac{\bnabla P}{\rhog} \! + \! \frac{\dtg}{\tau} (\bud \! - \! \bug) + \bFi (\bug) + \bFii, \label{eq:momentum_gas} \\
         \partial_{t} \bud + \bud \bcdot \bnabla \bud &= - \frac{1}{\tau} (\bud - \bug) + \bFi (\bud) + \bFii, \label{eq:momentum_dust}
    \end{align}
\end{subequations}
where ${ \dtg = \rhod / \rhog }$ is the dust-to-gas ratio and $\tau$ the dust's stopping time. Note that these equations are rigorously equivalent to those of paper I but easier to interpret.

\vspace{-1 \baselineskip}
\section{Background flow}
\label{sec:Background_flow}

\newcommand{\buK}{\bu_{\text{K}}} 
\newcommand{\hK}{h_{\text{K}}} 

The goal of paper I was to build analytical models of vortices embedded in \PPDs\ and permeated with dust. We started from Kida's gaseous vortex model \citep{Kida1981}, 
\begin{subequations}
    \label{eq:Kida_vortex_in_shearing_box}
    \begin{align}
        \buK &= \frac{S}{\alpha - 1} \left( \frac{Y}{\alpha} \, \eX - \alpha X \, \eY \right) , \label{eq:Kida_vortex_in_shearing_box_u} \\
        \hK &= \frac{S}{\alpha - 1} \bigg[ \left( \frac{S / 2}{\alpha - 1} - \Omega \right) X^{2} + \left( \frac{S / 2}{\alpha - 1} - \frac{\Omega}{\alpha} \right) Y^{2} \bigg] , \label{eq:Kida_vortex_in_shearing_box_P}
    \end{align}
\end{subequations}
and used an asymptotic expansion in small Stokes number to derive approximate dusty solutions ($\alpha$ is the vortex's aspect ratio, ${ h = P / \rhog }$ is a pseudo-enthalpy that conveniently replaces pressure, the Stokes number is ${ \St = \Omega \, \tau }$, and the subscript $\text{K}$ stands for \textit{Kida}, not \textit{Keplerian}).

We found that well-coupled particles spiral towards the core of anticyclones, but that this dust concentration process affects the vortices' evolution. However, on timescales shorter than ${ T_{\text{slow}} = 1 / (\tau \Delta \hK) }$, all our models collapsed to the same stationary leading-order-in-$\St$ flow,
\begin{subequations}
    \label{eq:dusty_vortex_model}
    \begin{align}
        \bug &= \buK , \label{eq:dusty_vortex_model_gas_velocity} \\
        \bud &= \buK + \tau \, \bnabla \hK , \label{eq:dusty_vortex_model_dust_velocity} \\
        \dtg &= \text{C}^{st.} . \label{eq:dusty_vortex_model_dust_to_gas_ratio}
    \end{align}
\end{subequations}
We will therefore use this solution, and assume that the phenomena we study occur on timescales shorter than~$T_{\text{slow}}$.

\vspace{-1 \baselineskip}
\section{Linear perturbation equations}
\label{sec:Linear_perturbation_equations}

To study the stability of our vortices, we follow the framework of linear perturbation theory, plus a few extra steps to deal with the spatial complexity inherent to vortices. Firstly, we write the general linear perturbation equations (\S\ref{sub:Linear_perturbation_equations_complete}). Then, we build an analog of the shearing box that follows a vortex streamline instead of a disc orbit (\S\ref{sec:VSB}). This `vortex shearing box' allows us to focus on small-wavelength perturbations and to simplify the perturbation equations accordingly (\S\ref{sub:Linear_perturbation_equations_small_wavelength_regime}). This enables a Fourier decomposition in space, thereby reducing the equations to a system of \ODEs\ in time (\S\ref{sub:Linear_perturbation_equations_Fourier_space}). Finally, we explain how we solve this system of equations (\S\ref{sec:LAVA}).

\vspace{-0.75 \baselineskip}
\subsection{The complete linear perturbation equations}
\label{sub:Linear_perturbation_equations_complete}

\newcommand{\eps}{\epsilon} 

Let us perturb the background flow, ${ f_{0} }$, with a perturbation,~${ \eps f_{1} }$, which we assume to be small, ${ 0 < \eps \ll 1}$. The total flow is then ${ f = f_{0} + \eps f_{1} }$, where $f$ represents any variable.

\newcommand{\bv}{\mathbf{v}} 

\newcommand{\dtgo}{\dtg_{0}} 
\newcommand{\fgo}{f_{g, 0}} 
\newcommand{\fdo}{f_{d, 0}} 
\newcommand{\bvo}{\bv_{0}} 

\newcommand{\ho}{h_{0}} 
\newcommand{\bugo}{\bu_{g, 0}} 
\newcommand{\budo}{\bu_{d, 0}} 
\newcommand{\rhodo}{\rho_{d, 0}} 

\newcommand{\hi}{h_{1}} 
\newcommand{\rhogi}{\rho_{g, 1}} 
\newcommand{\bugi}{\bu_{g, 1}} 
\newcommand{\rdi}{\varrho_{d, 1}} 
\newcommand{\rhodi}{\rho_{d, 1}} 
\newcommand{\budi}{\bu_{d, 1}} 

We inject this flow into Eqs.~\eqref{eq:Navier_Stockes_equations} but neglect the quadratic terms. This leads to the linear perturbation equations
\begin{subequations}
    \label{eq:linear_perturbation_equations}
    \begin{align}
        & \bnabla \bcdot \bugi = \,\, 0 , \label{eq:linear_pertubed_continuity_gas} \\
        & \partial_{t} \rdi + \rdi \bnabla \bcdot \budo + \budo \bcdot \bnabla \rdi = - \bnabla \bcdot \budi , \phantom{\frac{1}{1}} \label{eq:linear_pertubed_continuity_dust} \\
        & \begin{multlined}[t]
            \partial_{t} \bugi + \bugo \bcdot \bnabla \bugi + \bugi \bcdot \bnabla \bugo = - \bnabla \hi \! + \! \bFi (\bugi) \\
            + \frac{\dtgo}{\tau} (\budi - \bugi + \bvo \, \rdi) , \hspace{-0.3 cm}
        \end{multlined} \!\!\!\!\!\!\! \label{eq:linear_pertubed_momentum_gas} \\
        & \begin{multlined}[t]
            \partial_{t} \budi + \budo \bcdot \bnabla \budi + \budi \bcdot \bnabla \budo = \bFi (\budi) \\
            - \frac{1}{\tau} (\budi - \bugi) , \hspace{-1.5 cm}
        \end{multlined} \label{eq:linear_pertubed_momentum_dust}
    \end{align}
\end{subequations}
where ${ \rdi = \rhodi / \rhodo }$ is the relative perturbation in dust density and ${ \bvo = \budo - \bugo }$ is the background dust-to-gas drift. From this point onward, the subscript $0$ will be implicit in every instance of ${ \dtg }$. This lightens the notation without creating ambiguity.

Note that neglecting the quadratic terms leaves an error of size ${ \mathcal{O}(\eps^{2}) }$, and remember that the vortex model from Eq.~\eqref{eq:dusty_vortex_model} was only approximate, leaving another error of size ${ \mathcal{O}(\St) }$. These errors are acceptable as long as they dominated by the linear terms, \textit{i.e.} as long as ${ \St \ll \eps \ll 1 }$.

\vspace{-0.75 \baselineskip}
\subsection{The small-wavelength regime}
\label{sub:Linear_perturbation_equations_small_wavelength_regime}

The full linear perturbation equations are unfortunately quite hard to solve. Indeed, the space-dependent coefficients $\bugo$ and $\budo$ preclude any Fourier decomposition.\footnote{We provide more details on this point in \S\ref{sec:VSB}.} There are tools to solve eigenvalue problems whose unknowns are functions rather than vectors (see, \textit{e.g.}, \citealt{Burns+2020}), but they suffer from the same resolution and interpretability limitations as simulations. Therefore, we prefer to look for a regime in which Eqs.~\eqref{eq:linear_perturbation_equations} have a simpler spatial dependence.

\vspace{-0.75 \baselineskip}
\subsubsection{The `vortex shearing box'}
\label{ssub:Vortex_shearing_box}

Faced with a similar problem, \cite{RailtonPapaloizou2014} used a WKBJ approach. We prefer to emulate the shearing box idea, similarly to what \cite{OgilvieLatter2013} and \cite{OgilvieBarker2014} did for distorted discs. This approach is not only more transparent but also more versatile, since it could cover non-linear~perturbations.

Our strategy is to follow a geometric parcel as it moves along an elliptical streamline of the dust-free Kida flow from Eqs.~\eqref{eq:Kida_vortex_in_shearing_box_u}, and to only consider what happens in a small box around this reference parcel. 

\begin{figure}
    \centering
    \includegraphics[width = \linewidth]{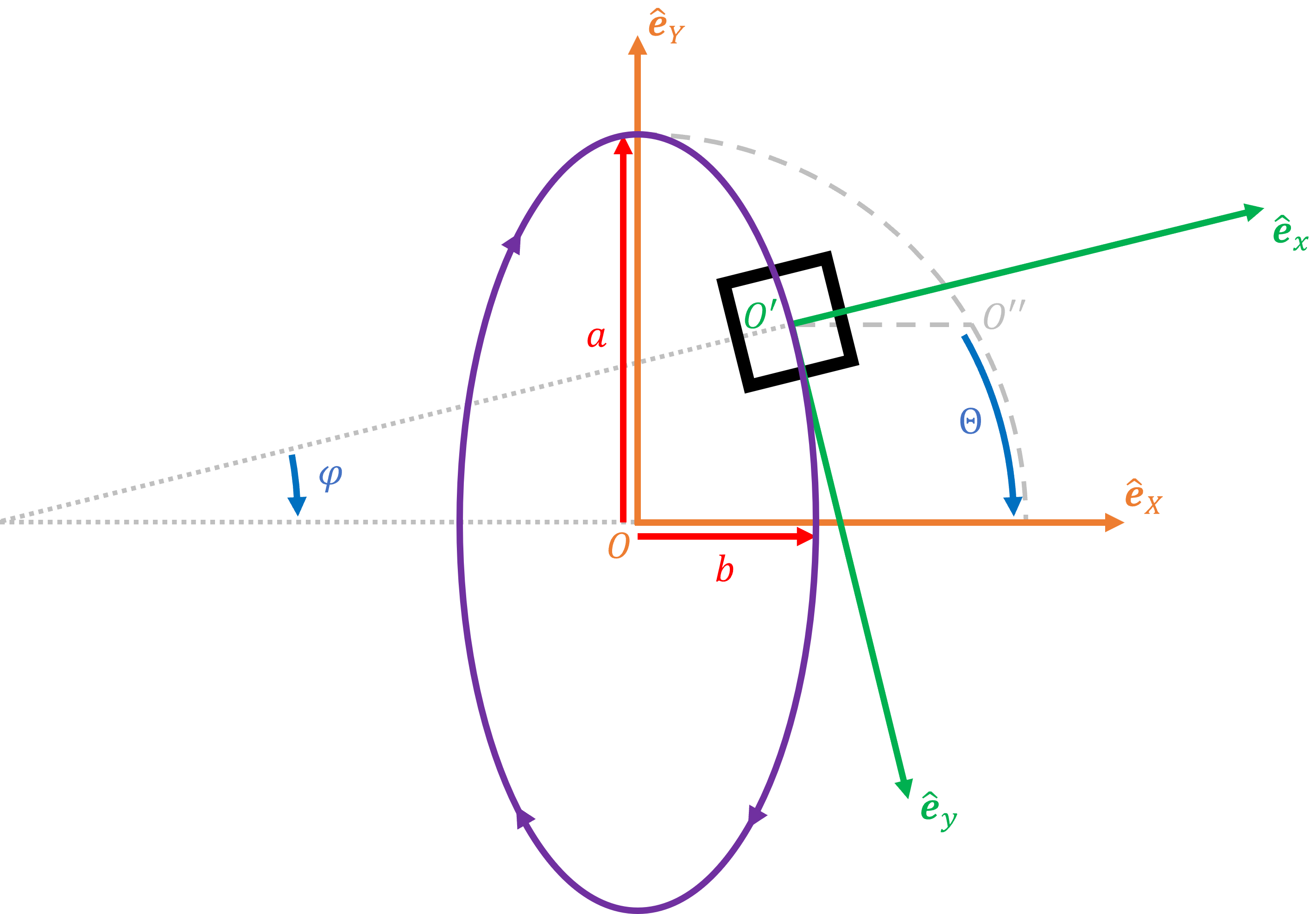}
    \caption{Schematic diagram of our `vortex shearing box'. $O$ is the centre of the standard shearing box, and $O'$ the centre of our box. The purple ellipse represents the reference streamline to which our box is attached. The red arrows measure this ellipse's semi-major axis $a$ and semi-minor axis $b$. In orange are the standard shearing box's unit vectors, and in green our box's unit vectors. The dashed grey lines show the geometrical construction of $\Theta$, and the dotted grey line the construction of $\varphi$. Those angles describe the instantaneous position and orientation of the box. Our box moves in the clockwise direction, so both angles increase over time.}
    \label{fig:Vortex_shearing_box_geometry}

    \vspace{-0.5 \baselineskip}

\end{figure}

The geometry of our `vortex shearing box' is described in \S\ref{sub:VSB_geometry} and Fig.~\ref{fig:Vortex_shearing_box_geometry}, but here is a short summary: the reference parcel is centered on ${ (X_{0}, Y_{0}) = (b \cos{\Theta}, - \alpha \, b \sin{\Theta}) }$, where $b$ is the semi-minor axis of the streamline and ${ \Theta = [S / (\alpha - 1)] \, t }$ is an angle. To denote position within the box, we use the Cartesian coordinates
\begin{subequations}
    \label{eq:VSB_change_of_coordinates_coordinates_bis}
    \begin{align}
        x &= + \cos{(\varphi)} (X - X_{0}) - \sin{(\varphi)} (Y - Y_{0}) , \label{eq:VSB_change_of_coordinates_coordinates_bis_x} \\
        y &= - \sin{(\varphi)} (X - X_{0}) - \cos{(\varphi)} (Y - Y_{0}) , \label{eq:VSB_change_of_coordinates_coordinates_bis_y} \\
        z &= - Z , \label{eq:VSB_change_of_coordinates_coordinates_bis_z}
    \end{align}
\end{subequations}
where $\varphi$ is another angle defined by ${ \alpha \tan{\varphi} = \tan{\Theta} }$. Since this is the angle between the $\ex$ and $\eX$ axes, it can be interpreted as the orientation of the box. $\Theta$, on the other hand, measures the box's position along the reference streamline.

\newcommand{\bA}{\mathbf{A}} 

In this box, the Kida flow~\eqref{eq:Kida_vortex_in_shearing_box_u} becomes
\begin{equation}
    \label{eq:Kida_vortex_in_VSB_u}
    \buK = S \, \frac{\alpha - \alpha^{-1}}{\alpha - 1} \left[ \frac{\sin{(2 \varphi)}}{2} (x \ex - y \ey) + \cos{(2 \varphi)} \, x \, \ey \right] \! .
\end{equation}
Note that contrary to the standard shearing box, this flow is not approximate but exact. Indeed, the Kida flow is linear in the coordinates, so there is no need to truncate a Taylor expansion at first order. Furthermore, $\buK$ being linear in the coordinates allows us to introduce a convenient matrix $\bA$ such that ${ \buK = \bA \, \bx }$. Finally, note that the standard shearing box's flow is a systematic shear, whereas our box's flow combines periodic shear and periodic strain. Figure~\ref{fig:Shear_and_Strain} gives an intuition for why that~is.

The change of coordinates also modifies our formulas~\eqref{eq:Kida_vortex_in_shearing_box_P}, \eqref{eq:forces_in_shearing_box_tidal_potential} and~\eqref{eq:forces_in_shearing_box_Coriolis} for the pressure, tidal and Coriolis forces from the shearing box. The new formulas are heavy and left to~\S\ref{sub:VSB_dynamics}.

Finally, the change of reference frame is non-Galilean so it brings a new set of fictitious forces: a second centrifugal force~${ \mathbf{f_{ce}^{\text{v}}} }$, a second Coriolis force~${ \mathbf{f_{Co}^{\text{v}}} }$, an Euler force~${ \mathbf{f_{Eu}^{\text{v}}} }$, and a force~$\mathbf{f_{co}^{\text{v-s}}}$ due to the composition of two changes of reference frame. Formulas for all these are provided in \S\ref{sub:VSB_dynamics}. One thing to note is that only the two Coriolis forces go into~$\bFi$, all other forces go into~$\bFii$ and are irrelevant to stability.

\subsubsection{Simplifications in the small-wavelength regime}
\label{ssub:Linear_perturbation_equations_small_wavelength_regime_simplifications}

If we consider a small-scale perturbation, we can study it over several wavelengths~$\lambda$ without leaving the box. Now if the box is small compared to the lengthscale on which a component of the background flow varies, it is safe to assume that this space-dependence has little impact on the evolution of our small-scale perturbation. This allows us to replace the background flow from Eq.~\eqref{eq:dusty_vortex_model} by a simpler, more uniform~flow.

\newcommand{\bvob}{\overline{\bv}_{0}} 

\newcommand{\Omegav}{\Omega_{\text{v}}} 

This argument can be made rigorous, as shown in \S\ref{sub:VSB_simplifications}. Remember that there are two types of space-dependent coefficients in Eqs.~\eqref{eq:linear_perturbation_equations}: those due to $\buK$, and those due to $\bvo$. We find that we can neglect the space-dependence of ${ \bvo }$ as long as $\lambda$ is small compared to $b$. This simplifies the linear perturbation equations to
\begin{small}
    \begin{subequations}
        \label{eq:linear_pertubation_equations_small_wavelength}
        \begin{align}
            & \bnabla \bcdot \bugi = 0, \label{eq:small_wavelength_equations_continuity_gas} \\
            & \! \Dt \rdi + \bvob \bcdot \bnabla \rdi = - \bnabla \bcdot \budi , \phantom{\frac{1}{1}} \label{eq:small_wavelength_equations_continuity_dust} \\
            & \! \Dt \bugi \! + \! \bugi \! \bcdot \! \bnabla \buK \! = \! - \bnabla \hi \! + \! \bFi (\bugi) \! + \! \frac{\dtgo}{\tau} (\budi \! - \! \bugi \! + \! \bvob \, \rdi) , \nonumber \refstepcounter{equation} \\
            & \! \Dt \budi \! + \! \bvob \bcdot \bnabla \budi \! + \! \budi \bcdot \bnabla \buK \! = \! \bFi (\budi) \! - \! \frac{1}{\tau} (\budi \! - \! \bugi) , \!\!\! \label{eq:small_wavelength_equations_momentum_dust}
        \end{align}
    \end{subequations}
\end{small}
\!\!where ${ \Dt = \partial_{t} + \buK \bcdot \bnabla }$ is the Eulerian derivative and
\begin{multline}
    \label{eq:space_independent_part_of_the_dust_drift}
    \bvob = \tau \bigg\{ \big[ q_{1} X_{0} \cos(\varphi) - q_{2} Y_{0} \sin (\varphi) \big] \ex \\
    - \big[ q_{1} X_{0} \sin (\varphi) + q_{2} Y_{0} \cos (\varphi) \big] \ey \bigg\}
\end{multline} 
is the value taken by the dust's drift $\bvo$ in the centre of the vortex shearing box. ${ q_{1} \! = \! \Omegav (\Omegav / 4 \! - \! \Omega) }$ and ${ q_{2} \! = \! \Omegav (\Omegav / 4 \! - \! \Omega / \alpha) }$ are scalar constants, and ${\Omegav \! = \! 2 S / (\alpha \! - \! 1)}$ is the vortex's \textbf{half}-turnover frequency.

Equations~\eqref{eq:linear_pertubation_equations_small_wavelength} may not seem much nicer than Eqs.~\eqref{eq:linear_perturbation_equations}, but almost all the space-dependent coefficients have disappeared, the only ones left being in $\buK$.

\subsection{Simplifications in Fourier space}
\label{sub:Linear_perturbation_equations_Fourier_space}

\newcommand{\bK}{\mathbf{K}} 

The simplifications from \S\ref{ssub:Linear_perturbation_equations_small_wavelength_regime_simplifications} enable us to decompose the variables on a time-explicit Fourier basis, \smash{${ f_{1}(t, \bx) = \tilde{f}_{1}(t) \, \e^{i \bk(t) \cdot \bx} }$}. The time-dependency of the wave-vector introduces a new type of space-dependent coefficients, of the form ${ \rd_{t} \bk \bcdot \bx }$. Fortunately, these can cancel the preexisting ${ \buK (\bx) \bcdot \bk }$ coefficients, provided that the wave-vector ${ \bk }$ evolves according to
\begin{equation}
    \label{eq:wavevector_evolution}
    \rd_{t} \bk = - \bA^{\mathrm{T}} \,\, \bk .
\end{equation}
Note that we are not making any assumption about the physics here, we are simply choosing a convenient Fourier basis. If we define the dimensionless wavenumbers $K_x$, $K_y$ and $K_z$ such that
\begin{equation}
    b \bk(\varphi = \frac{\pi}{2}) = K_x \, \ex + K_y \, \ey + K_z \, \ez , \nonumber
\end{equation}
we find that the solutions to Eq.~\eqref{eq:wavevector_evolution} take the form
\begin{multline}
    \bK (\varphi) = \frac{2 \alpha^{2} \, K_{x} - (\alpha^{2} - 1) \, \sin (2 \varphi) \, K_{y}}{2 \alpha \, \sqrt{\cos^{2} (\varphi) + \alpha^{2} \sin^{2} (\varphi)}} \, \ex \\
    + \frac{\sqrt{\cos^{2} (\varphi) + \alpha^{2} \, \sin^{2} (\varphi)}}{\alpha} \, K_{y} \, \ey + K_z \, \ez . \label{eq:wavevector_definition}
\end{multline}

\newcommand{\bOmega}{\mathbf{\Omega}} 

\newcommand{\thi}{\tilde{h}_{1}} 
\newcommand{\tbugi}{\tilde{\bu}_{g, 1}} 
\newcommand{\trdi}{\tilde{\varrho}_{d, 1}} 
\newcommand{\tbudi}{\tilde{\bu}_{d, 1}} 

Now that all the undesirable terms have been removed, the perturbation equations \eqref{eq:linear_pertubation_equations_small_wavelength} become
\begin{small}
    \begin{subequations}
        \label{eq:linear_perturbation_equations_Fourier_transformed}
        \begin{align}
            & i \bk \bcdot \tbugi = \,\, 0 , \label{eq:Fourier_transformed_equations_continuity_gas} \\
            & \rd_{t} \trdi + i \left( \bk \bcdot \bvob \right) \, \trdi = - i \bk \bcdot \tbudi , \label{eq:Fourier_transformed_equations_continuity_dust} \\ 
            & \rd_{t} \tbugi \! + \! \bA \, \tbugi \! = \! - i \bk \, \thi \! - \! 2 \bOmega \, \tbugi \! + \! \frac{\dtg}{\tau} (\tbudi \! - \! \tbugi \! + \! \bvob \, \trdi) , \!\!\! \label{eq:Fourier_transformed_equations_momentum_gas} \\
            & \rd_{t} \tbudi \! + \! i \left( \bk \bcdot \bvob \right) \, \tbudi \! + \! \bA \, \tbudi \! = \! - 2 \bOmega \, \tbudi \! - \! \frac{1}{\tau} (\tbudi \! - \! \tbugi) , \label{eq:Fourier_transformed_equations_momentum_dust}
        \end{align}
    \end{subequations}
\end{small}
\!\!where ${ \bOmega }$ is the time-dependent matrix giving ${ \bFi (\mathbf{f}) \! = \! - 2 \bOmega \, \mathbf{f} }$. 

This system is controlled by seven dimensionless parameters: the dust-to-gas ratio $\dtg$, the Stokes number of the particles $\St$, the vortex's aspect ratio $\alpha$, the disc's shear rate ${S/\Omega}$, and the Fourier mode's wavenumbers ${ \{K_{x}, \, K_{y}, \, K_{z}\} }$.

The good news is that Eqs.~\eqref{eq:linear_perturbation_equations_Fourier_transformed} form a system of \ODEs, whereas Eqs.~\eqref{eq:linear_pertubation_equations_small_wavelength} formed a system of \PDEs. The bad news is that the standard \SI's equations were already too complex to be solved analytically. Ours are made even worse by the time dependence of~$\bk$, $\bvob$, $\bA$ and~$\bOmega$, so we can only solve them approximately in asymptotic regimes, and numerically in the rest of parameter space. A description of our numerical solver is given in \S\ref{sec:LAVA}.

\vspace{-0.5 \baselineskip}
\section{The waves}
\label{sec:Waves}

The vortex simulations of \cite{Fu+2014}, \cite{Raettig+2015}, \textit{etc.} suggest that the dust's backreaction triggers an instability akin to the \SI. Now remember that the \SI\ is due to the interaction between an inertial wave and a dust density wave \citep{SquireHopkins2020, Magnan2024b}. Therefore, before solving Eqs.~\eqref{eq:linear_perturbation_equations_Fourier_transformed}, we should study what waves propagate in vortices. To do so, we focus on the test particle regime,~${ \dtg = 0 }$, where the dust does not affect the gas. We start with the gas waves in \S\ref{sub:Waves_gas}, then move on to the dust waves in \S\ref{sub:Waves_dust}.

\vspace{-0.5 \baselineskip}
\subsection{Behaviour of the gas}
\label{sub:Waves_gas}

The gas is unaffected by the dust, so we can study it in isolation. Gas perturbations follow the simplified equations
\begin{subequations}
    \label{eq:Fourier_transformed_equations_gas_only}
    \begin{align}
        i \bk \bcdot \tbugi &= \,\, 0 , \label{eq:Fourier_transformed_equations_gas_only_continuity} \\
        \rd_{t} \tbugi + \bA \, \tbugi &= - i \bk \, \thi - 2 \bOmega \, \tbugi . \label{eq:Fourier_transformed_equations_gas_only_momentum}
    \end{align}
\end{subequations}
Note that since we did not neglect the space-dependent part of any gas coefficient, those equations are exact. 

The continuity equation~\eqref{eq:Fourier_transformed_equations_gas_only_continuity} slaves the vertical velocity perturbation to the horizontal velocity perurbations, and pressure acts as a Lagrange multiplier enforcing mass conservation. This means that the system, even though it has four variables, only supports two modes.

\cite{LesurPapaloizou2009} worked on this Floquet problem. They showed that the characteristic multipliers can only depend on three parameters: ${ S / \Omega }$, ${ \alpha }$, and the wavevector's latitude~${ \theta = \arctan{(K_x / K_z)} }$. \cite{LebovitzZweibel2004} considered the symmetries of the evolution matrix, and found that either the gas supports two waves of similar nature and opposite frequency, or the system is unstable.

\vspace{-0.5 \baselineskip}
\subsubsection{Inertial waves}
\label{ssub:Inertial_waves}

\newcommand{\bg}{\mathbf{g}} 

Since the Coriolis force is the only available restoring force, the two waves must be inertial waves. But due to the time-dependency of the vortex shearing box's rotation rate and thereby of the vortex-induced Coriolis force, those inertial waves are not simply sinusoidal. This forces us to use Floquet rather than Fourier theory. Within this framework, the waves take the form
\begin{equation}
    \label{eq:Floquet_mode}
    \tilde{f}_{1} (t) = F_{1} (\varphi) \, \e^{(\gamma_{\text{iw}} + i \omega_{\text{iw}} ) t} ,
\end{equation}
where ${ F_{1} }$ is a ${ 2 \pi }$-periodic function, ${ \e^{(\gamma_{\text{iw}}  + i \omega_{\text{iw}} ) T_{v} } }$ is called the characteristic multiplier of the mode and ${ s = \gamma_{\text{iw}}  + i \omega_{\text{iw}}  }$ the Floquet exponent of the mode. We shall furthermore call $\gamma_{\text{iw}} $ the Floquet growth rate and $\omega_{\text{iw}} $ the Floquet frequency. 

Note that $\omega_{\text{iw}} $ is only defined modulo $\Omegav$. To break this degeneracy, we define the Floquet frequency as the only representative of the congruence class that lives between ${- \Omegav /2 }$ and ${+ \Omegav / 2 }$. Lebowitz and Zweibel's result implies that the Floquet frequencies of the two inertial waves are opposite.

\newcommand{\tUgy}{\tilde{U}_{g, y}} 

In the 2D axisymmetric case, \textit{i.e.} when ${ K_{z} = K_{y} = 0 }$, the equations are tractable analytically. Indeed, the radial velocity must be null to avoid compressiblity. Eq.~\eqref{eq:Fourier_transformed_equations_gas_only_momentum} then shows that the azimuthal velocity responds to advection, and that pressure takes whatever value it needs to compensate the Coriolis force radially. This leads to
\begin{subequations}
    \label{eq:zonal_flow}
    \begin{align}
        \thi &=  2 i \, (\Omega - \rd_{t} \varphi) \, \frac{\cos^{2} (\varphi) + \alpha^{2} \sin^{2} (\varphi)}{\alpha} \times \frac{b \, \tUgy}{K_{x}} , \label{eq:zonal_flow_P} \\
        \tbugi (t) &= \tUgy \, \sqrt{\cos^{2} (\varphi) + \alpha^{2} \sin^{2} (\varphi)^{2}}\, \ey , \label{eq:zonal_flow_u}
    \end{align}
\end{subequations}
where $\tUgy$ is a constant quantifying the mode's amplitude.

Since these oscillations induce a purely azimuthal velocity perturbation and fail to propagate (their Floquet frequency is null), we call them \textit{zonal flows}. To lighten notation, we write ${ \tbugi (t) \! = \! \tUgy \, F_{\text{zf}} (\varphi) \, \ey }$ and ${ \omega_{\text{zf}} = 0 }$.

\begin{figure}
    \centering
    \includegraphics[width = \linewidth]{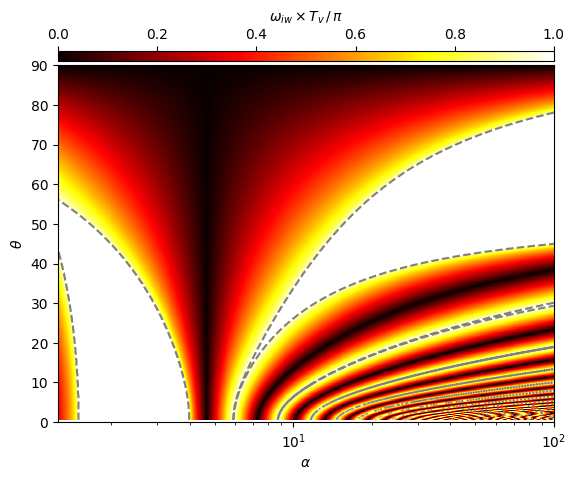}
    \caption{
        Frequency ${ \omega_{\text{iw}} }$ of the inertial waves propagating in the core of Kida's vortex, as a function of the vortex's aspect ratio~${ \alpha }$ and of the wavevector's latitude~${ \theta }$. The regions surrounded by dashed grey lines are elliptically unstable. Their being white means that ${ \omega_{\text{iw}} }$ is congruent to the vortex's turnover~frequency ${ S / (\alpha \! - \! 1) }$. Our analysis of the dust-induced instabilities in \S\ref{sec:Instability} will focus on the elliptically stable band that runs from ${ \alpha = 4}$ to ${ \alpha = 6 }$. \\ \textit{Parameters:}~${ S / \Omega = 1.5 }$.
    }
    \label{fig:gas_frequencies}
\end{figure}

Unfortunately, in all other cases, it becomes difficult to compute the Floquet exponent analytically. We can however compute it numerically, as per Fig.~\ref{fig:gas_frequencies}.

\vspace{-0.5 \baselineskip}
\subsubsection{The elliptical instability}
\label{ssub:Elliptical_instability}

The gas system is unstable whenever an inertial wave has a Floquet frequency congruent to the vortex's \textbf{full} turnover frequency. This leads to an unstable resonance called the \EI. The white regions surrounded by dashed grey lines in Fig.~\ref{fig:gas_frequencies} show that the core of most protoplanetary vortices are subject to this instability, except those of aspect ratio between 4 and 6. The growth rates are plotted in Fig.~\ref{fig:Lesur_Papaloizou}.

\subsection{Behaviour of the dust}
\label{sub:Waves_dust}

We can study the dust modes by setting the gas perturbations to zero. The perturbation equations~\eqref{eq:linear_perturbation_equations_Fourier_transformed} simplify to
\begin{subequations}
    \label{eq:linear_perturbation_equations_Fourier_transformed_dust_only}
    \begin{align}
        \rd_{t} \trdi + i \, ( \bk \bcdot \bvob ) \, \trdi &= - i \bk \bcdot \tbudi , \label{eq:Fourier_transformed_equations_continuity_dust_only} \\ 
        \rd_{t} \tbudi + i \, ( \bk \bcdot \bvob ) \, \tbudi + \bA \, \tbudi &= - 2 \bOmega \, \tbudi - \tbudi / \tau . \label{eq:Fourier_transformed_equations_momentum_dust_only}
    \end{align}
\end{subequations}
This system is of order four, so the dust supports four modes.

\subsubsection{Dust density waves}
\label{ssub:Dust_density_waves}

\newcommand{\tRd}{\tilde{R}_{d}} 

The structure of Eqs.~\eqref{eq:linear_perturbation_equations_Fourier_transformed_dust_only} permits a perturbation in dust density alone, governed by the first-order \ODE
\begin{equation}
    \label{eq:dust_density_wave_equation}
    \rd_{t} \trdi + i \, ( \bk \bcdot \bvob ) \, \trdi = 0 , 
\end{equation}
whose solution is
\begin{equation}
    \label{eq:dust_density_wave}
    \trdi (t) = \tRd \, \e^{ -i \int_{0}^{t} \rd s \,\, \left[ \bk (s) \cdot \bvob (s) \right] } ,
\end{equation}
where $\tRd$ is a constant indicating the mode's amplitude. 

Since dust density perturbations are simply advected by the background dust drift, we call this mode the \textit{dust density wave} and use the compressed notation ${ \trdi (t) = \tRd \, F_{\text{ddw}} (t) }$.

Due to the time-dependency of the background dust drift in the vortex shearing box, this wave is non-modal. Eq.~\eqref{eq:dust_density_wave} shows that one of the Floquet frequencies is 
\begin{equation}
    \label{eq:dust_density_wave_frequency}
    \omega_{\text{ddw}} = - \frac{\int_{0}^{T_{v}} \rd t \,\, \left[ \bk (t) \cdot \bvob (t) \right]}{T_{v}} .
\end{equation}
It may not be between ${ - \Omegav / 2 }$ and ${ + \Omegav / 2 }$, but it is easy to write and evaluate, so we shall call it `the' Floquet frequency of the dust density wave. In the axisymmetric case, \textit{i.e.} when ${ K_{y} = 0 }$, this frequency can be computed analytically: 
\begin{equation}
    \label{eq:dust_density_wave_frequency_develloped}
    \frac{\omega_{\text{ddw}}}{\Omega} = - \St \, K_{x} \times \left( \frac{S}{\Omega} \right) \frac{1 + \left( \frac{S}{\Omega} \right) \alpha - \alpha^{2}}{(\alpha - 1)^{2}} .
\end{equation}

\subsubsection{Other dust waves}
\label{ssub:Waves_dust_other}

The other three dust waves are perturbations in dust velocity governed by Eq.~\eqref{eq:Fourier_transformed_equations_momentum_dust_only}. They are harder to study rigorously, but in the regime of small particles, we expect them to be strongly damped by the ${ - \tbudi / \tau }$ term. Therefore, we do not expect those waves to play any significant role.

\vspace{-0.5 \baselineskip}
\section{The instabilities}
\label{sec:Instability}

If there is enough dust to affect the gas, the vortex becomes unstable. We start in~\S\ref{sub:2D_instability} with the 2D axisymmetric version of this instability. We present the raw results from \nameofmycode~(\S\ref{ssub:2D_instability_numerical_results}) and show that the instability is a form of \SI~(\S\ref{ssub:2D_instability_numerical_nature}). Then, we move on to the 3D axisymmetric instability in \S\ref{sub:3D_instability} and to the non-axisymmetric instability in~\S\ref{sub:Non-axisymmetric_instability}.

\subsection{The 2D axisymmetric instability}
\label{sub:2D_instability}

Restricting the analysis to 2D is interesting in two regards. First, it reduces the number of variables, which helps push the analysis further. Second, the dusty vortex simulations of \cite{Fu+2014}, \cite{Raettig+2015}, \cite{Surville+2016}, \cite{SurvilleMayer2019} and \cite{Lovascio+2022} were all 2D and all showed an unknown instability, whose nature it would be good to establish.

Assuming axisymmetry is useful for the same reasons, but may be less physically motivated. We shall discuss this in \S\ref{sec:Discussion}.

\vspace{-0.5 \baselineskip}
\subsubsection{Numerical results}
\label{ssub:2D_instability_numerical_results}

\begin{figure}
    \centering
    \includegraphics[width = \linewidth]{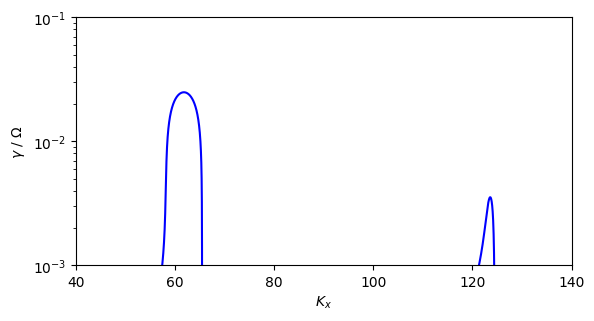}
    \caption{
        Growth rate of the 2D axisymmetric perturbations ${ \gamma }$ as a function of the adimensional `radial' wavenumber $K_{x}$, in the regime of dilute and well-coupled dust. \\ \textit{Parameters:}~${ S / \Omega = 1.5 }$, ${ \dtg = 0.1 }$, ${ \St = 0.01 }$, ${ \alpha = 4.5 }$, ${ K_{y} \! = \! K_{x} \! = \! 0 }$.
    }
    \label{fig:2D_instability}
\end{figure}

Fig.~\ref{fig:2D_instability} shows that when ${ \dtg > 0 }$, there are two bands of instability. Fig.~\ref{fig:2D_eigenmode} shows the structure of the fastest-growing mode.

Crucially, this fastest-growing mode has a large wavenumber, validating \textit{a posteriori} the usage of the vortex box. Furthermore, since the \EI\ is inactive in 2D \citep{LesurPapaloizou2009}, our instability must be due to the dust's backreaction. We shall discuss later, in \S\ref{sec:Discussion}, whether this is the unknown instability of \cite{Fu+2014}, \cite{Raettig+2015}, \textit{etc.}

\vspace{-0.5 \baselineskip}
\subsubsection{Analysis of the instability's nature}
\label{ssub:2D_instability_numerical_nature}

\begin{figure}
    \centering
    \includegraphics[width = \linewidth]{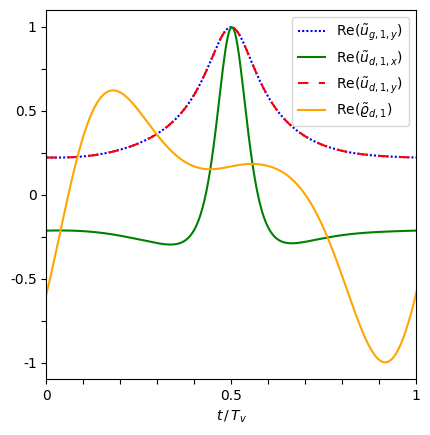}
    \caption{
        Fastest-growing eigenmode of the 2D axisymmetric instability. The modulii are normalised to one. Since the particles are well-coupled, the blue lines corresponding to ${ \tilde{u}_{g, 1, y} }$ and the red lines corresponding to ${ \tilde{u}_{d, 1, y} }$ are almost identical. \\ \textit{Parameters:}~Same as in Fig.~\ref{fig:2D_instability}, except that ${ K_{x} = 60 }$.
    }
    \label{fig:2D_eigenmode}
\end{figure}

We suspect that our instability is an \RDI. Our intuition is based on the fact that (i)~the instability relies on the dust's backreaction onto the gas; (ii)~it appears in thin bands, hinting at a resonance; (iii)~the second band appears at twice the horizontal frequency of the first band, suggesting a fundamental and its harmonic, and thus that the resonance involves waves; (iv)~the gas velocity perturbation resembles a zonal flow (compare the blue line of Fig.~\ref{fig:2D_eigenmode} to Eq.~\ref{eq:zonal_flow_u}).

To confirm this rigorously, we need to generalise the \RDI\ theories of \cite{SquireHopkins2018b} and \cite{Magnan2024b}. Indeed, the waves described in \S\ref{sec:Waves} are not simple sine waves, so the standard theory is inapplicable.

This generalisation is quite involved, so we leave it to \S\ref{sec:Non_modal_RDI_theory}. We find that a non-modal gas wave can resonate with a non-modal dust wave if their Floquet frequencies $\omega_{\text{g}}$ and $\omega_{\text{d}}$ satisfy
\begin{equation}
    \label{eq:resonance_condition}
    \omega_{\text{g}} = \omega_{\text{d}} + \frac{2 n \pi}{T_{v}}, \,\,\, n \in \mathbb{Z} .
\end{equation}
This is less rigid than the classical \RDI\ criterion, which demands that the dust and gas Fourier frequencies be identical.

\begin{figure}
    \centering
    \includegraphics[width = 0.95 \linewidth]{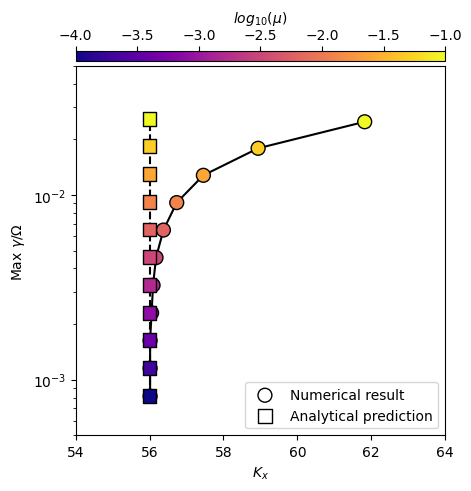}
    \caption{
        Comparison between the wavenumber $K_x$ and growth rate $\gamma$ of the fastest-growing 2D axisymmetric mode, as predicted by \nameofmycode\ (circles) versus Eqs.~\eqref{eq:resonance_condition}~and~\eqref{eq:growth_rate_squared_2D} (squares). A cell's colour indicate the value of its dust-to-gas ratio $\dtg$. \\ \textit{Parameters:}~Same as in Fig.~\ref{fig:2D_instability}, except that $\dtg$ varies and that $K_{x}$ is optimised to maximise the growth rate.
    }
    \label{fig:Comparison_RDI_theory_vs_LAVA_2D}
\end{figure}

Applying this result to the zonal flows from \S\ref{sub:Waves_gas} and the dust density waves from \S\ref{sub:Waves_dust}, we can evaluate the resonant `radial' wavenumber,
\begin{equation}
    \nonumber
    K_x = \frac{2 n \, (\alpha - 1)}{\St \, (\alpha^{2} - [S / \Omega] \, \alpha - 1)} .
\end{equation}
Furthermore, if the resonance condition is verified, we can make a semi-analytical prediction for the growth rate of the resulting instability (Eq.~\ref{eq:growth_rate_squared_2D}). Figure~\ref{fig:Comparison_RDI_theory_vs_LAVA_2D} shows that this prediction is in excellent agreement with the \nameofmycode\ simulations, especially in the \mbox{low-$\dtg$} regime where our asymptotic theory was derived.

This confirms that the instability of Fig.~\ref{fig:2D_instability} is an \RDI\ based on the vortex's zonal flows. That makes it a close cousin of the \SI, which relies on inertial waves.

\subsubsection{Why the \SI\ extends to 2D in vortices}
\label{ssub:2D_instability_why}

In discs, an \RDI\ based on zonal flows is impossible, because they do not propagate, so they cannot resonate with the dust's radial drift. Zonal \!flows \!still \!fail \!to \!propagate \!in vortices, so one may wonder what makes resonance possible there.

\newcommand{\tudix}{\tilde{u}_{d, 1, x}} 
\newcommand{\tugiy}{\tilde{u}_{g, 1, y}} 

To understand this, let us consider the regime of test particles once more. This is the cleanest way to investigate the effect of a gas wave on the dust (the `forward action' of \citealt{Magnan2024b}). We show in \S\ref{sec:Response_of_test_particles_to_zonal_flows} that if the gas wave is the zonal flow from Eq.~\eqref{eq:zonal_flow}, the dust velocity is approximately 
\begin{align}
    \label{eq:dust_forced_by_zonal_flow_velocity}
    \tudix \approx 2 \tau \left[ \left(\rd_{t} \varphi\right) - \Omega \right] \tUgy \, F_{\text{zf}} .
\end{align}

Setting aside mathematical rigor for a second, this equation represents the well-known tendency of dust to migrate towards pressure bumps. Indeed, Eq.~\eqref{eq:dust_forced_by_zonal_flow_velocity} can be abusively reframed as ${ \bud = 2 \tau (\rd_{t} \varphi - \Omega) \, \bug }$. Furthermore, in 2D axisymmetric settings, the radial component of the gas momentum equation~\eqref{eq:Fourier_transformed_equations_gas_only_momentum} gives ${ i k_{x} \, \thi = 2 (\rd_{t} \varphi - \Omega) \, \tugiy }$, which can be abusively reframed as ${ \bnabla h = 2 (\rd_{t} \varphi - \Omega) \, \bug }$. Combining those two approximations leads to ${ \bud = \tau \bnabla h }$, confirming that dust follows pressure gradients.

The key difference between vortices and discs is that in elongated vortices, the Coriolis parameter \smash{${ \rd_{t} \varphi - \Omega }$} can change sign. Indeed, \smash{${ \rd_{t} \varphi = \frac{S}{\alpha (\alpha - 1)} \left[ \cos^{2} (\varphi) + \alpha^{2} \sin^{2} (\varphi) \right] }$} and \smash{${ S \approx \frac{3}{2} \Omega }$}, so if ${ \alpha > (1 + \sqrt{7}) / 2 }$, the parameter switches sign as the box travels from major axis to minor axis. This is crucial, because it changes the sign of the pressure perturbation $\thi$.

\begin{figure}
    \centering
    \includegraphics[width = \linewidth]{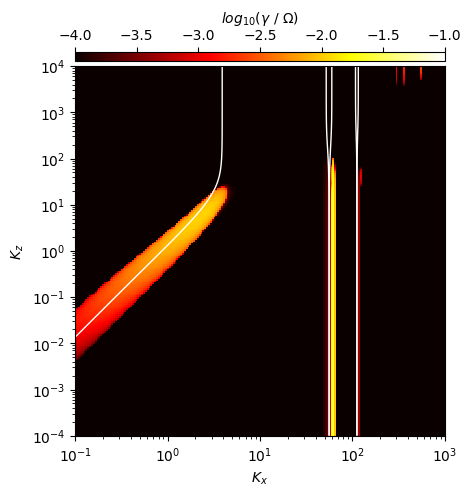}
    \caption{
        Growth rate ${ \gamma }$ of the 3D perturbations as a function of the `radial' and vertical wavenumbers ${ K_{x} }$ and ${ K_{z} }$ in the regime of dilute and well-coupled dust, ${ \St, \dtg \ll 1}$. The white lines represent the modes selected by the resonance condition~\eqref{eq:resonance_condition}. \\ \textit{Parameters:}~Same as in Fig.~\ref{fig:2D_instability}, except that $K_{z}$ is non-zero.
    }
    \label{fig:Comparison_RDI_theory_vs_LAVA_3D}
\end{figure}

\begin{figure*}
    \centering

    \begin{minipage}{\textwidth}
        \centering
        \includegraphics[width = \linewidth]{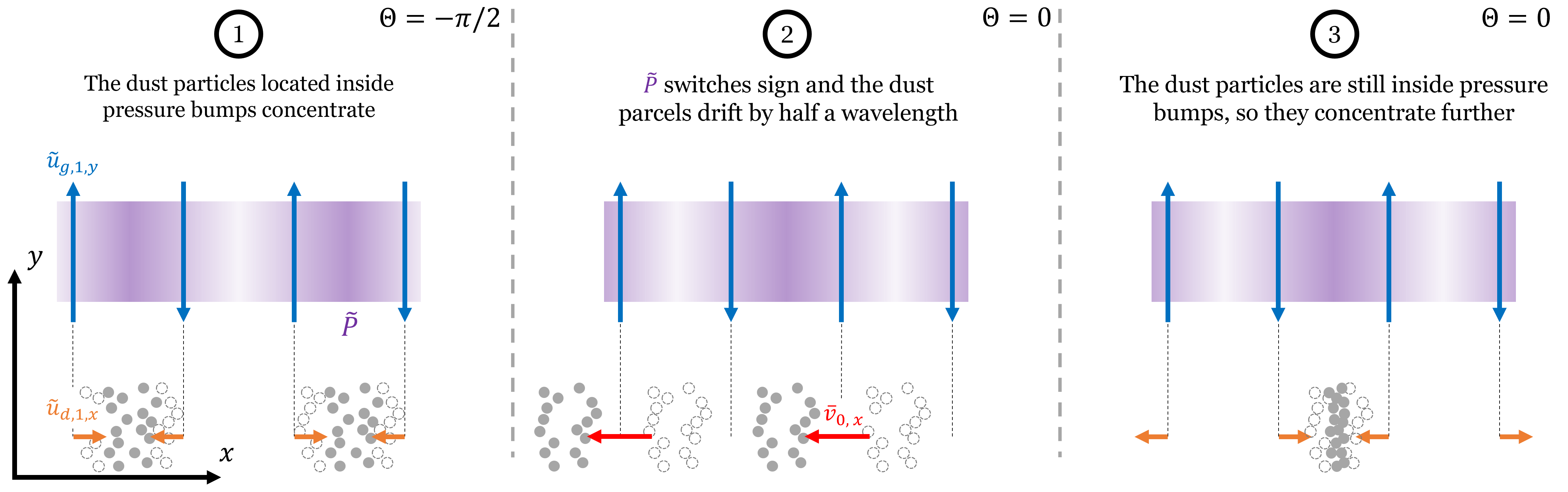}
        \caption{
            Comic strip explaining how the vortex \SI's forward action proceeds in 2D. The first panel is drawn when the box is at major axis ($\Theta = -\pi/2$), and the last two panels when the box is at minor axis ($\Theta = 0$). The top part of each panel shows a zonal-flow mode: blue denotes the gas velocity perturbations and purple the gas pressure perturbations. The bottom part of each panel shows how dust responds to this gas mode: orange denotes the (radial) dust velocity perturbations, and some dust particles appear in grey. They get closer and closer to each other, meaning that the dust density increases in those regions. The red arrows depict the dust's radial drift.
        }
        \label{fig:Forward_action}
    \end{minipage}


    \begin{minipage}{\textwidth}
        \centering
        \includegraphics[width = \linewidth]{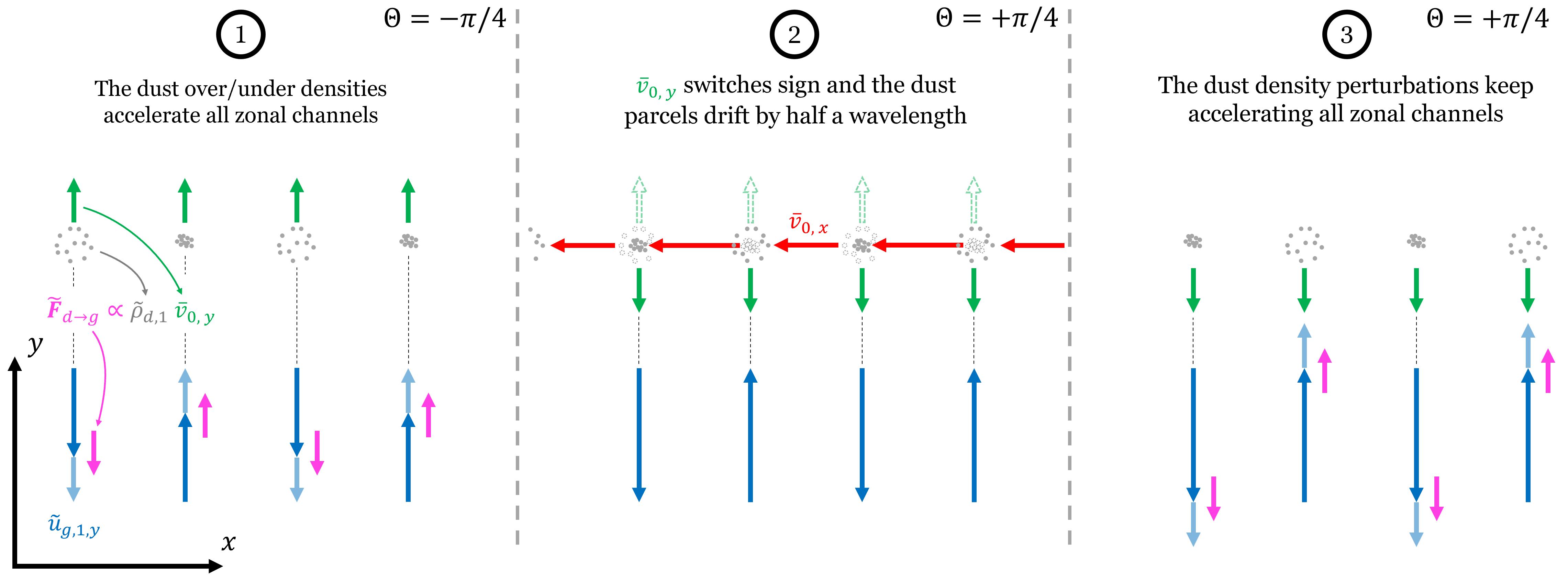}
        \caption{
            Comic strip explaining how the vortex \SI's backward reaction proceeds in 2D. The first panel is drawn when the box is halfway between major and minor axis (${\Theta = - \pi/4}$), and the last two panels when the box is halfway between minor and major axis (${\Theta = +\pi/4}$). The top part of each panel shows several over-dense and under-dense dust fluid parcels (in grey), as well as the direction of their azimuthal drift (in green). The bottom part of each panel shows the backreaction force anomaly exerted by those dust parcels on the gas (in pink), and how that force modifies the amplitude of the gas' zonal flow (in blue).
        }
        \label{fig:Backward_reaction}
    \end{minipage}

    \vspace{-1 \baselineskip}

\end{figure*}

Figure~\ref{fig:Forward_action} explains how this alternating pressure drives the growth of dust density perturbations. Essentially, we just saw that the dust density in a parcel increases/decreases when that parcel is in a pressure bump/trough (panel 1). Now at any given time, half of the zonal flow channels are pressure bumps and half are pressure troughs. Therefore, one could think that dust parcels spend as much time in pressure bumps and pressure troughs as they drift radially. But if the dust takes exactly one vortex half-turnover time to drift by one zonal flow wavelength,\footnote{\label{footnote_1}Note that this is what the resonance condition~\eqref{eq:resonance_condition} means.} then pressure bumps and troughs exchange positions just as the dust leaves a channel to enter the next one (panel 2). Therefore, the parcel is always inside a pressure bump, and its dust density can increase without limit (panel 3). This explains why the forward action of the vortex \SI\ survives in 2D.

Regarding the backward reaction (\textit{i.e.} the process by which a dust density wave amplifies zonal flows), it survives in 2D for similar reasons. Specifically, an overly dense dust parcel exerts an excess backreaction on the gas, in the direction of the background dust drift $\bvob$ (panel 1 of Fig.~\ref{fig:Backward_reaction}). Since the drifting dust parcel visits zonal channels of alternating direction, one could think that it spends as much time accelerating the zonal flow as it does slowing it down. But Eq.~\eqref{eq:space_independent_part_of_the_dust_drift} shows that the azimuthal component of the background dust drift~$\overline{v}_{0, y}$ changes sign as the box travels from major axis to minor axis (panel 2). So if the dust parcel crosses from one zonal channel to the next at the same time that $\overline{v}_{0, y}$ changes sign,\footnoteref{footnote_1} then the parcel always accelerates the zonal flow (panel~3).

For readers familiar with our previous papers on \RDI\ theory \citep{Magnan2024a, Magnan2024b}, we provide some mathematical backing for this tentative physical picture in \S\ref{sec:Non_modal_RDI_theory}.

\vspace{-1 \baselineskip}
\subsection{The 3D axisymmetric instability}
\label{sub:3D_instability}

Let us now consider 3D perturbations. The extra degrees of freedom may activate new unstable bands or strengthen the 2D ones. Studying this may help interpret the 3D dusty vortex simulations of \cite{Lyra+2018} and \cite{Raettig+2021}. 

Once again, we study the stability of the dusty vortex of aspect ratio ${ \alpha = 4.5 }$ because it is elliptically stable \citep{LesurPapaloizou2009}. Figure~\ref{fig:Comparison_RDI_theory_vs_LAVA_3D} presents the growth rate against $K_{x}$ and $K_{z}$. Once again, we find that when ${ \dtg > 0 }$, the dust's backreaction triggers an instability.

\enlargethispage{+0.5 \baselineskip}
The structure of Fig.~\ref{fig:Comparison_RDI_theory_vs_LAVA_3D} is quite complex. There is a diagonal band at low ${ K_{x} }$, two vertical bands at large ${ K_{x} }$, and several weak and narrow bands at large ${ K_{z} }$ (the `pimples').

\begin{figure*}
    \centering
    \includegraphics[width = \linewidth]{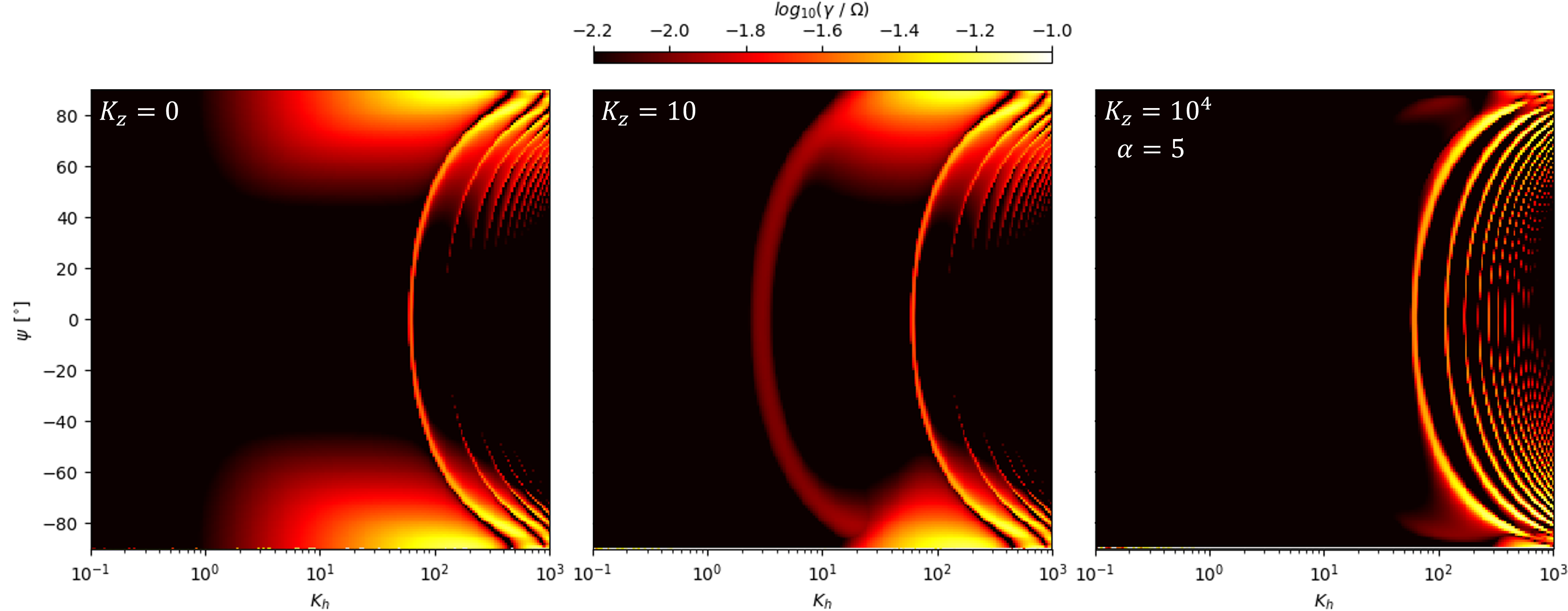}
    \caption{
        Growth rate ${ \gamma }$ of the non-axisymmetric perturbations as a function of the `horizontal' wavenumber ${ K_{h} }$ and of the wavevector's longitude ${ \psi }$. $K_{z}$ is chosen such that the left plot relates to the vertical bands of Fig.~\ref{fig:Comparison_RDI_theory_vs_LAVA_3D}, the central plot to the diagonal band, and the right plot to the pimples. \textit{Parameters:}~Same as in Fig.~\ref{fig:Comparison_RDI_theory_vs_LAVA_3D}, except that $K_{y}$ is non-zero. Additionally, we use ${ \alpha = 5 }$ in the third panel to make the pimples more apparent.
        }
    \label{fig:Effect_of_psi}
\end{figure*}

The vertical bands are easy to interpret. Indeed, they connect to the bands of Fig.~\ref{fig:2D_instability} when $K_{z}$ becomes small. This indicates that the vertical bands represent the same `zonal flow \RDI' as the 2D instability. Of course, when ${ K_{z} \neq 0 }$ the vortex supports inertial waves rather than zonal flows. But when ${ K_z \ll K_x }$, the difference between inertial waves and zonal flows is marginal. In particular, the Floquet frequency ${ \omega_{\text{iw}} }$ is non-zero, but negligible compared to $\Omega$ or $\Omegav$. It only becomes significant when ${ K_z \sim K_x }$, and indeed this is approximately where the vertical bands disappear.

\enlargethispage{+0.5 \baselineskip}
We expect the diagonal band to be even more closely related to the standard \SI. Indeed, it has the same slope of two in the $\log{(K_{z})}$-$\log{(K_{x})}$ plane \citep{YoudinGoodman2005}. To confirm our prediction, we determine the values of $K_{x}$ and $K_{z}$ for which the dust density wave and one of the inertial waves satisfy the resonance condition~\eqref{eq:resonance_condition},\footnote{${ \omega_{\text{ddw}} }$ is given by Eq.~\eqref{eq:dust_density_wave_frequency_develloped}, but ${ \omega_{\text{iw}} }$ is computed numerically.} and we overlay the selected modes upon Fig.~\ref{fig:Comparison_RDI_theory_vs_LAVA_3D}. This shows that all the growing modes are resonant, confirming that the diagonal band and the pimples are `inertial wave \RDIs' akin to the \SI. The diagonal band corresponds to ${n = 0}$ and the pimples to ${ n \geq 1 }$.

\subsection{The non-axisymmetric instability}
\label{sub:Non-axisymmetric_instability}

Non-axisymmetric perturbations are irrelevant for the standard \SI\ because non-axisymmetric structures are sheared out by the disc's differential rotation. But in Kida's vortex, the net mean shear is null, so those modes might live and grow. 

Figure~\ref{fig:Effect_of_psi} presents the results of our experiments. We varied the longitude of the mode's wavevector, ${ \psi = \arctan{(K_{y} / K_{x})} }$, and the horizontal wavenumber, \smash{${ K_{h} = \sqrt{K_{x}^{2} + K_{y}^{2}} }$}.

The left panel shows the 2D non-axisymmetric instability. When ${\psi = 0}$ we recover the main band from Fig.~\ref{fig:2D_instability}. Not only does it extend to non-axisymmetric settings, its growth rate increases. The second band is weaker but also strengthens with longitude, so it only appears near the top and the bottom of the present figure. Interestingly, several new bands appear at high $ |\psi| $. This is in good agreement with Eq.~\eqref{eq:resonance_condition}, which allows a band for every integer $n$. The subtlety is that the second band disappears from the present figure at ${ \psi = 0 }$ because it becomes weak, whereas the higher-order bands disappear because they become inactive.

There is also a smooth cloud of unstable modes at extreme longitudes and small scales. It does not look resonant, but we could not establish its nature. That being said, we think that in real protoplanetary vortices -- whose vorticity profiles are non-Kida and therefore induce persistent shear -- those modes would be short-lived. We are therefore wary of over-interpreting them.

The central panel is in 3D. We find exactly the same features as in 2D, plus a new band at lower $K_{h}$ corresponding to the diagonal band of Fig.~\ref{fig:Comparison_RDI_theory_vs_LAVA_3D}. It extends to high longitudes while keeping its growth rate $\gamma$ and radial wavenumber $K_{x}$ constant. This suggests that $K_{y}$ does not play much of a role.

The last panel focuses on the pimples. They are stronger and easier to detect when~${ \alpha = 5 }$ than when~${ \alpha = 4.5 }$, hence why we chose this value. Just like the diagonal and vertical bands, they extend to non-axisymmetric settings without much change to~$K_{x}$. The growth rate increases marginally.

\section{Exploration of parameter space}
\label{sec:Exploration_of_parameter_space}

Exploring parameter space will help us solidify our identification of the instability as an \RDI, understand how it manifests itself, determine which vortices are most stable, predict in which part of the disc vortices are most capable of catalysing planetesimal formation, \textit{etc.}

The shear rate ${ S/\Omega }$ will not vary significantly from ${ 3/2 }$ in \PPDs, and we already studied the role of $\bK$ in \S\ref{sub:Non-axisymmetric_instability}. That leaves us with three parameters to study: $\dtg$, $\St$ and $\alpha$. In the upcoming subsections, we consider them one at a time.

\vspace{-1 \baselineskip}
\subsection{Impact of the dust-to-gas ratio}
\label{sub:Exploration_of_parameter_space_dtg}

\begin{figure}
    \centering
    \includegraphics[width = \linewidth]{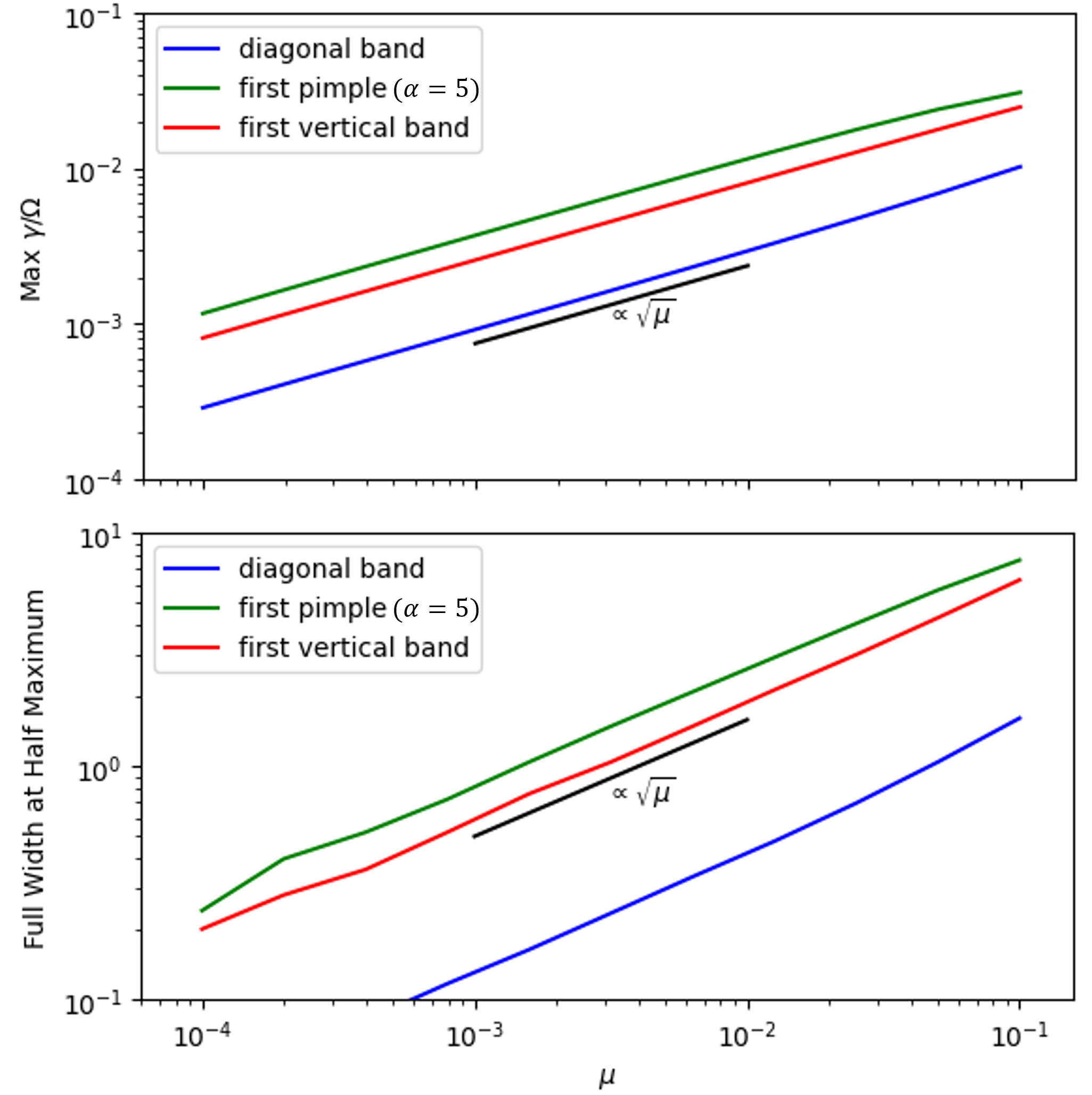}
    \caption{
        \textit{Top:} Maximum growth rates in the different bands of Fig.~\ref{fig:Comparison_RDI_theory_vs_LAVA_3D} as a function of the dust-to-gas ratio~$\dtg$.
        \textit{Bottom:} Same thing, but for the full width at half maximum of the bands. \\ \textit{Parameters:}~The slices ranges from ${ K_{x} = 1 }$ to ${ K_{x} = 10 }$ for the diagonal band and from ${ K_{x} = 40 }$ to ${ K_{x} = 80 }$ for the first vertical band and the first pimple. As for $K_{z}$, we use $10$ for the diagonal band, $0$ for the first vertical band and $10^{4}$ for the first pimple. Additionally, we use ${ \alpha = 5 }$ for the first pimple.
    }
    \label{fig:Effect_of_low_mu}
\end{figure}

Figure~\ref{fig:Effect_of_low_mu} shows how ${ K_{z} = \text{C}^{\text{st.}} }$ slices of the three main bands respond to the dust-to-gas ratio. 

The first plot shows that the maximum growth rate in any given band scales with ${ \sqrt{\dtg} }$. This is the behaviour expected of an \RDI\ \citep{SquireHopkins2018b}. The second plot presents the evolution of the bands' widths, measured at half maximum and given in units of $K_{x}$. They also scale with ${ \sqrt{\dtg} }$, in accordance with \RDI\ detuning theory \citep{Magnan2024a}.

A crucial caveat is that those scalings are only valid in gas-dominated regime, \textit{i.e.} when ${ \dtg < 1 }$. Indeed, the gas-dominated and dust-dominated \SIs\ are known to be different instabilities \citep{YoudinGoodman2005, SquireHopkins2020}. We expect something similar to happen in vortices.

\vspace{-1 \baselineskip}
\subsection{Impact of the Stokes number}
\label{sub:Exploration_of_parameter_space_St}

Similarly, we investigate how ${ \St \, K_{z} = \text{C}^{\text{st.}} }$ slices of the different bands evolve with the Stokes number~${ \St }$. 

This scaling is justified by the fact that ${ \omega_{\text{iw}} }$ is a function of ${ \theta = \arctan{[(\St \, K_{x}) / (\St \, K_{z})]} }$ while ${ \omega_{\text{ddw}} }$ a function of ${ \St \, K_{x} }$. Therefore, Eq.~\eqref{eq:resonance_condition} is really a condition on ${ \St \, \bK }$, not ${ \bK }$.

One consequence is that as ${ \St }$ get smaller, so does the wavelength of the fastest growing mode. Therefore, the small-scale assumption from \S\ref{sub:Linear_perturbation_equations_small_wavelength_regime} must be valid for small enough dust grains. However, if the wavelength becomes too small, turbulent viscosity becomes important.

\begin{figure}
    \centering
    \includegraphics[width = \linewidth]{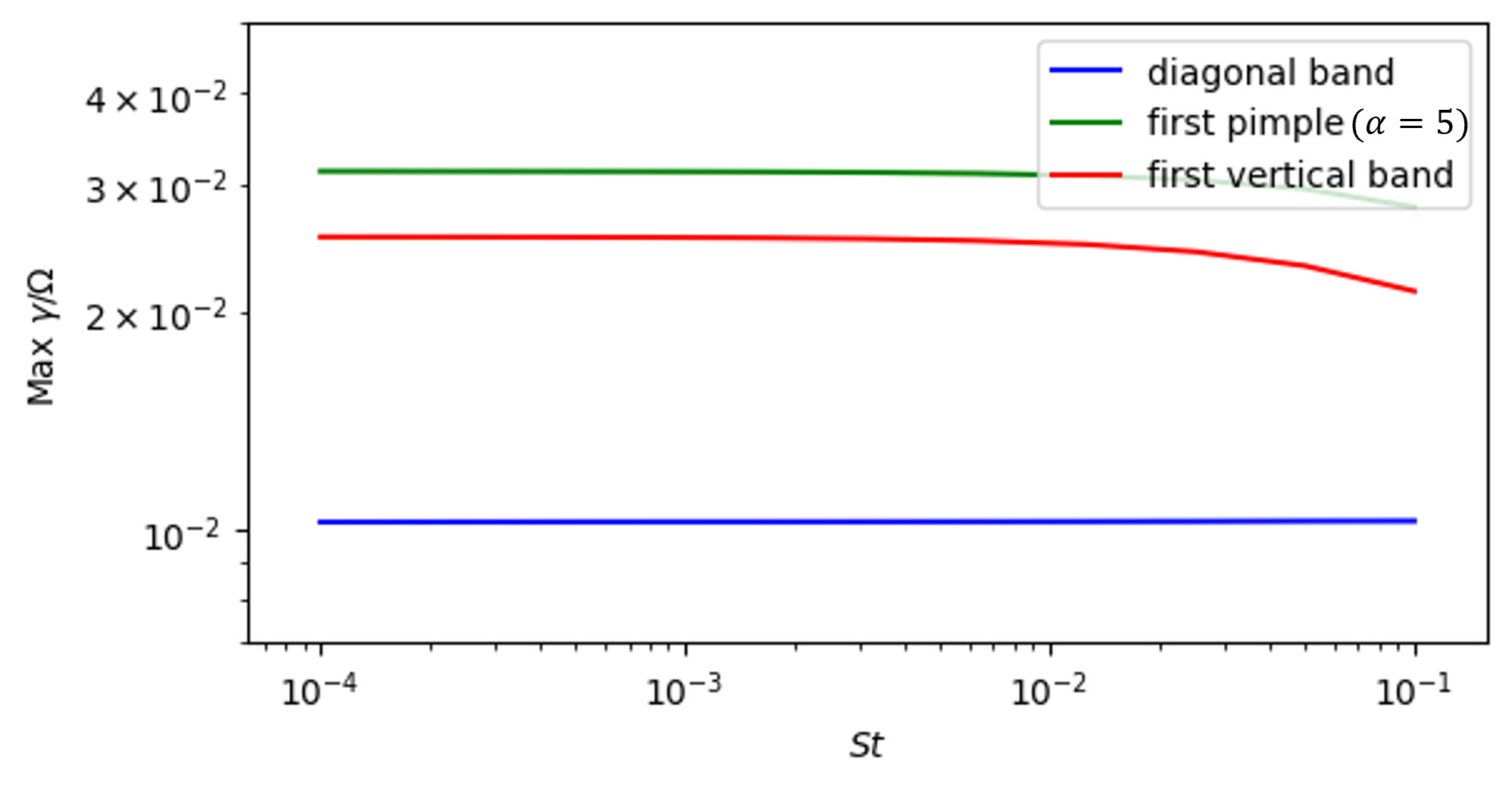}
    \caption{
        Maximum growth rate in the different bands of Fig.~\ref{fig:Comparison_RDI_theory_vs_LAVA_3D} as a function of $\St$, in the limit of small particles. \\ \textit{Parameters:}~the slices ranges from ${ \St \, K_{x} = 0.01 }$ to ${ \St \, K_{x} = 0.1 }$ for the diagonal band and from ${ \St \, K_{x} = 0.4 }$ to ${ \St \, K_{x} = 0.8 }$ for the first vertical band and the first pimple. As for ${ \St K_{z} }$, we use $0.1$ for the diagonal band, $0$ for the first vertical band and $10^{2}$ for the first pimple. Additionally, we use ${ \alpha = 5 }$ for the first pimple.
    }
    \label{fig:Effect_of_St}
\end{figure}

Figure~\ref{fig:Effect_of_St} shows that the growth rate of the vortex \SI\ is independent of $\St$. This is surprising because the growth rate of the standard \SI\ scales with $\St$ in the limit of small particles \citep{SquireHopkins2020, Magnan2024b}.

We can explain this by referring to our previous paper on the \SI\ \citep{Magnan2024b}. The difference is that in discs, the azimuthal component of the background dust drift scales with $\St^{2}$ \citep{Nakagawa+1986}, whereas in vortices it scales with $\St$ (\textit{cf.} Eq.~\ref{eq:space_independent_part_of_the_dust_drift}). This makes the `slow' azimuthal mechanism of the \SI's backward reaction just as fast as the `fast' radial mechanism. Then, just like in the standard \SI, the `slow' backward mechanism couples with the `fast' trapping-in-pressure-bumps forward mechanism to close a positive feedback loop leading to instability. But since one leg of the loop is stronger than usual by a factor ${ 1/\St }$, the whole instability becomes stronger. The `slow forward - fast backward' loop still exists, but becomes subdominant.

\subsection{Impact of the vortex aspect ratio}
\label{sub:Exploration_of_parameter_space_alpha}

The last parameter is the vortex's aspect ratio ${ \alpha }$. We first describe what happens in the elliptically stable band (\S\ref{ssub:sub:Exploration_of_parameter_space_alpha_elliptically_stable_band}), then describe how the vortex \SI\ interacts with the \EI\ (\S\ref{ssub:sub:Exploration_of_parameter_space_alpha_interaction_with_EI})

\subsubsection{In the elliptically stable band}
\label{ssub:sub:Exploration_of_parameter_space_alpha_elliptically_stable_band}

\begin{figure*}
    \centering
    
    \begin{minipage}{\textwidth}
        \centering
        \includegraphics[width = \linewidth]{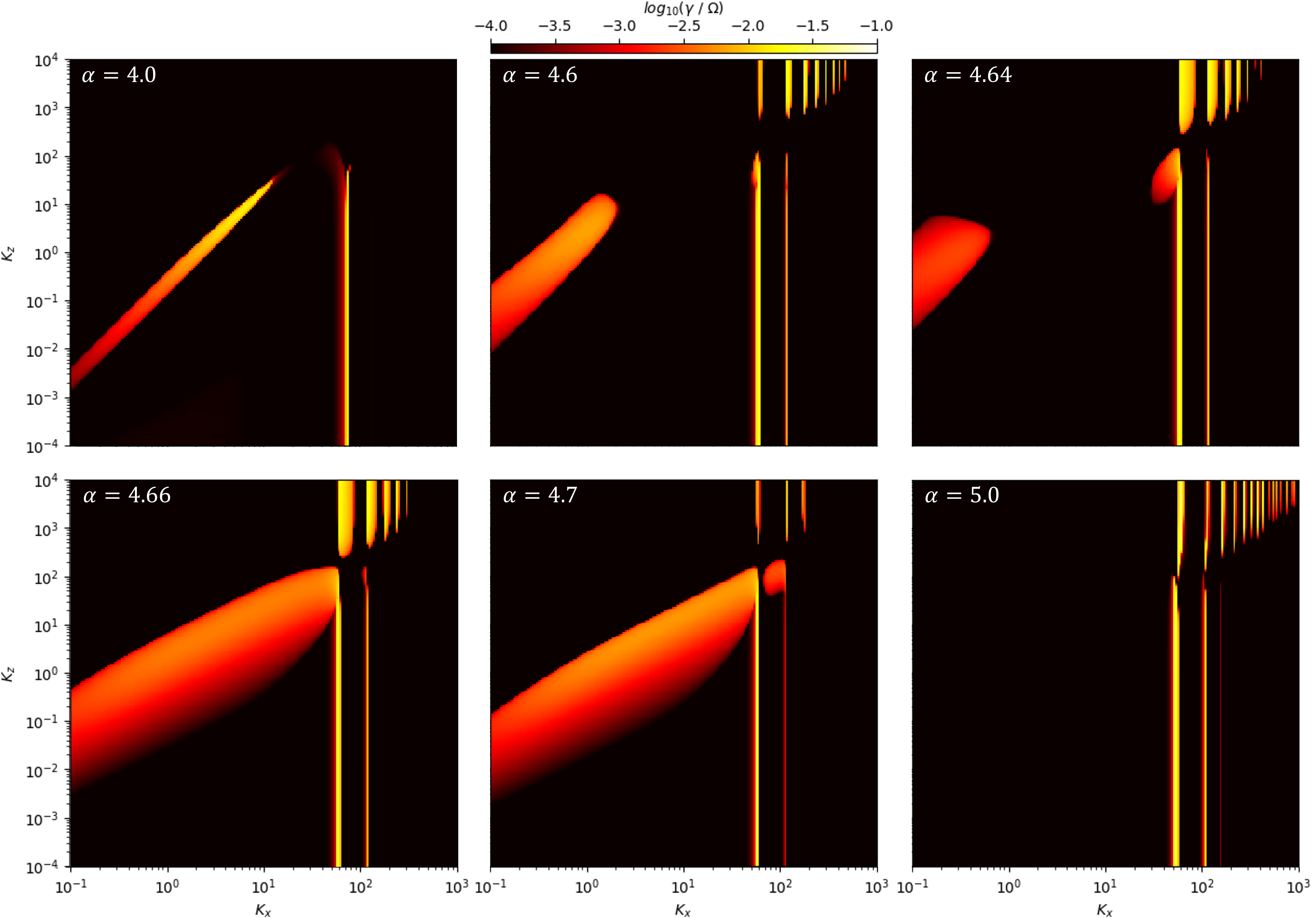}
        \caption{
            Same as Fig.~\ref{fig:Comparison_RDI_theory_vs_LAVA_3D}, but for six different values of the vortex's aspect ratio $\alpha$, all in the elliptically stable band.
        }
        \label{fig:Effect_of_alpha_elliptically_stable_band}
    \end{minipage}

    \vspace{+2 \baselineskip}

    \begin{minipage}{\textwidth}
        \centering
        \includegraphics[width = \linewidth]{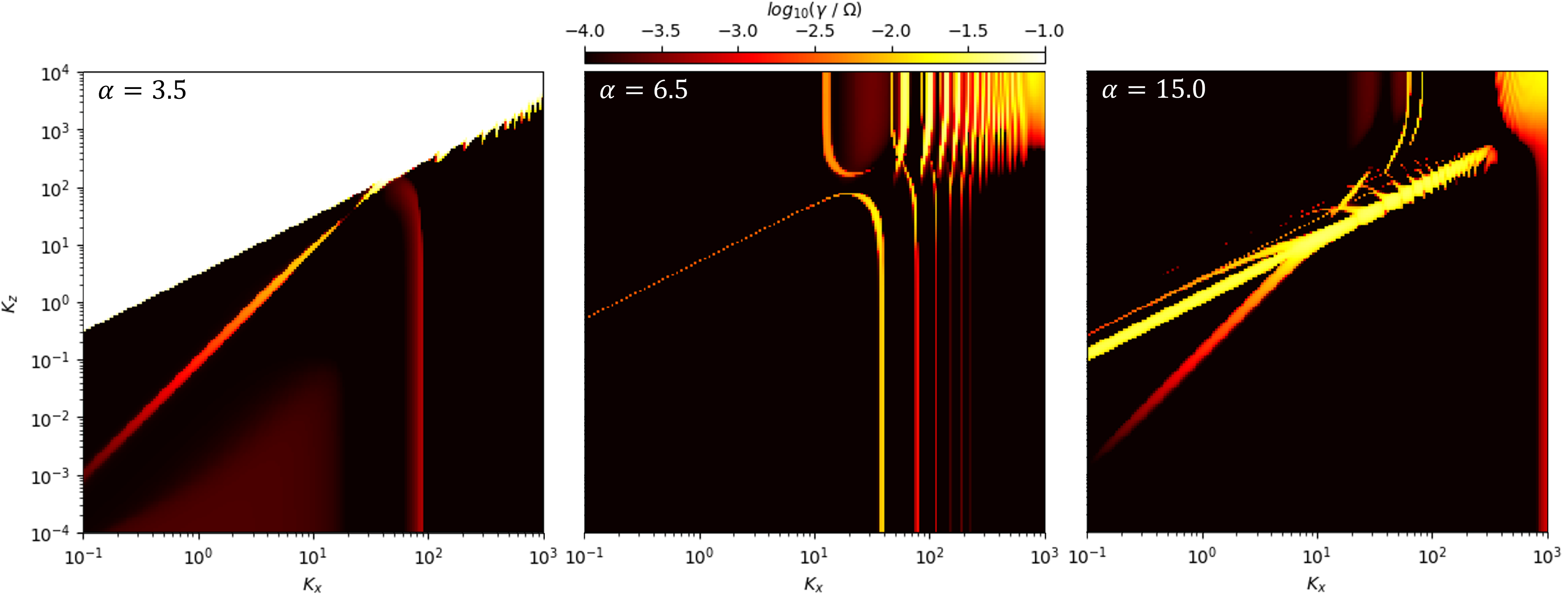}
        \caption{
            Same as Fig.~\ref{fig:Effect_of_alpha_elliptically_stable_band}, but for three values of $\alpha$ where the \EI\ is active.
        }
        \label{fig:Effect_of_alpha_interaction_with_EI}
    \end{minipage}

\end{figure*}

Fig.~\ref{fig:Effect_of_alpha_elliptically_stable_band} shows how Fig.~\ref{fig:Comparison_RDI_theory_vs_LAVA_3D} evolves as ${ \alpha }$ increases from~4 to~6. The \EI\ is inactive in this band, so any instability must be the `vortex \SI'. Its dependence on $\alpha$ is quite complex: between ${ \alpha = 4.0 }$ and ${ \alpha = 4.5 }$ the second vertical band and the pimples appear; between 4.5 and 4.6 the pimples become much stronger; between 4.6 and 4.64 the diagonal band disappears and the pimples broaden; between 4.64 and 4.66 the diagonal band reappears with a different slope; between 4.66 and 4.7 the pimples become much weaker; and finally between 4.7 and 5.0 the diagonal band disappears again while the pimples become much stronger and numerous.

We do not have an explanation for this complex behavior, but we note that the first vertical band remains dominant at all points between ${ \alpha = 4 }$ and ${ \alpha = 6 }$, so the leading-order evolutions of all those vortices are probably quite similar.

\subsubsection{Interaction with the elliptical instability}
\label{ssub:sub:Exploration_of_parameter_space_alpha_interaction_with_EI}

Let us now investigate how the vortex \SI\ interacts with the \EI. To do so, we consider three more aspect ratios in Fig.~\ref{fig:Effect_of_alpha_interaction_with_EI}.

The ${ \alpha = 3.5 }$ vortex shows a strongly unstable area at low latitude (\textit{i.e.} when ${ K_{z} \gg K_{x} }$). This is the centrifugal instability of \cite{LesurPapaloizou2009}. At higher latitudes, we see the diagonal and vertical bands of the vortex \SI, but they are much weaker and as such would not play any role on the vortex's demise. The dust's effect on the growth rate of the centrifugal instability seems negligible.

The last two vortices are subject to the parametric~\EI\ \citep{LesurPapaloizou2009}. Usually, it is independent of wavelength but strongly dependent on $\theta$, so it appears in log-log space as a thin band of slope one. We see that band in the middle pannel, but only up to ${ K_{x} = 30 }$. Beyond this point, it gets hijacked by the first vertical band of the vortex \SI. This means that dust makes the \EI\ wavelength-dependent: it behaves normally at large wavelengths, but the dust quenches it at small wavelengths. The middle panel also shows that at ${ \alpha = 6.5 }$, the diagonal band of the vortex \SI\ is still absent, but that there are many vertical bands and pimples.

The last panel shows that as $\alpha$ increases, the \EI\ becomes stronger so dust quenching retreats to shorter wavelengths. It also shows the return of the vortex \SI's diagonal band, the disappearance of the pimples and of the vertical bands, and the appearance of ancillary bands around the \EI\ band. Just like in \S\ref{ssub:sub:Exploration_of_parameter_space_alpha_elliptically_stable_band}, this complex behavior is hard to interpret.

\section{Discussion}
\label{sec:Discussion}

We have shown that the \SI\ is active in vortices, and explored its dependence on several parameters. Let us now discuss the validity of our results (\S\ref{sub:Discussion_due_diligence}), how they compare to simulations (\S\ref{sub:Discussion_comparison_to_simulations}), and what they mean for planet formation theory, vortex observations, vortex evolution, and \RDI\ theory (\S\ref{sub:Discussion_applications}).

\subsection{Due diligence}
\label{sub:Discussion_due_diligence}

Analytical results are only as good as the worst of the assumptions supporting them. We should therefore defend our choices. We start in \S\ref{ssub:Due_diligence_relevance_of_hypotheses} by showing that our hypotheses are astrophysically relevant, then we verify in \S\ref{ssub:Due_diligence_self-consistency_of_hypotheses} that they are self-consistent.

\subsubsection{Relevance of the hypotheses}
\label{ssub:Due_diligence_relevance_of_hypotheses}

Our first assumption is that the gas is incompressible. This would be an issue if the instability grew quickly compared to the time it takes a sound wave to cross the vortex, but it does not. Indeed, if the vortex's size is comparable to the disc's scale height, the sound crossing time is just one orbit.

Our second assumption was to fix the Stokes number $\St$ rather than the particle size $a$. In the Epstein regime of drag, $\St$ scales with ${ a / \rho_{g} }$ \citep{ChiangYoudin2010}. Proto-planetary vortices exhibit slightly higher gas densities in their core \citep{SurvilleBarge2015}, so strictly speaking $\St$ decreases as a particle drifts towards the vortex's centre. But particles drift slowly, so this variation happens on a longer timescale than the instability, and we can safely neglect it.

We further assumed that all particles have the same size. This is questionable. Indeed, the poly-disperse \SI\ behaves differently from the mono-disperse \SI\ \citep{Krapp+2019, Paardekooper+2020, Paardekooper+2021, McNally+2021}. Fortunately, vortices preferentially capture particles of a certain size \citep{BargeSommeria1995}. Therefore, we expect narrower particle size distributions in vortices than in discs. And since the dust's drift speed is proportional to $\St$, we expect even stronger size segregation in the vortex's core.

We chose to work in the regime of dilute dust. There are arguments for and against that regime. The average dust-to-gas mass ratio in \PPDs\ is of order 0.01, so if vortices start with the same composition, Fig.~\eqref{fig:Effect_of_low_mu} allow us to determine at what age vortices develop a virulent instability. However, \cite{Miranda+2017} argue that if the vortex is created by the \RWI, then the pressure bump that caused the \RWI\ already contained a lot of dust, which is concentrated further by the merger of several small vortices into one large-scale vortex. Consequently, they find an initial dust-to-gas ratio of order~10 in their vortex.

Other questionable points include the small size of our particles and our modelling the dust as a pressure-less fluid. We presented cases for and against those assumptions in paper~I.

Ultimately, we think that our weakest hypothesis is neglecting the vertical component of gravity, because that filters out vertical shear, buoyancy waves, and dust sedimentation.

\subsubsection{Self-consistency of the hypotheses}
\label{ssub:Due_diligence_self-consistency_of_hypotheses}

To keep the linear stability analysis manageable, we neglected the slow evolution of the background vortices. Consequently, our results are only valid on short timescales. Specifically, if $\gamma$ is the growth rate of the instability and ${ \Omega / (\St \, |\Delta \hK|) }$ the vortex evolution timescale, we need
\begin{equation}
    \label{eq:self_consistency_condition}
    \frac{\gamma}{\Omega} \gg 2 \pi \, \St \, \frac{|\Delta \hK|}{\Omega^{2}} .
\end{equation}
The left-hand side of Eq.~\eqref{eq:self_consistency_condition} scales with ${ \dtg^{1/2} \, \St^{0} }$ while the right-hand side with ${ \dtg^{0} \, \St^{1} }$, so there must exist a critical Stokes number scaling with $\dtg^{1/2}$ below which our hypotheses are self-consistent.

\begin{figure}
    \centering
    \includegraphics[width = \linewidth]{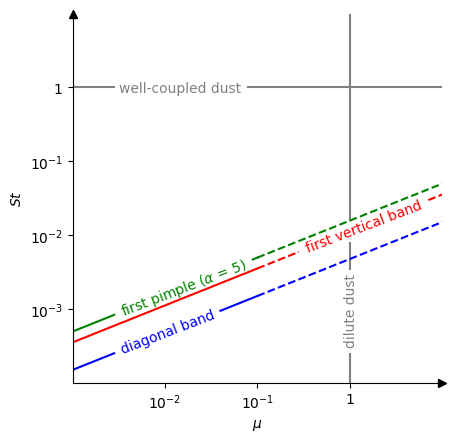}
    \caption{
        Diagram presenting the regions of parameter space where the self-consistency condition~\eqref{eq:self_consistency_condition} is satisfied (assuming a factor 10 between the left-hand side and the right-hand side). The dashed lines indicate that we are extrapolating our scaling laws beyond the range of dust-to-gas ratios explored in Fig.~\ref{fig:Effect_of_low_mu}.
    }
    \label{fig:Due_diligence}
\end{figure}

Figure~\ref{fig:Due_diligence} shows that for an interstellar dust-to-gas ratio of~1\% (and if we interpret $\gg$ as a factor 10), the particles would have to be very small (${ \St < 10^{-3} }$). In fact, even once the vortex has concentrated dust to the point where ${ \dtg \approx 1 }$, we still need ${ \St < 10^{-2} }$.

This sounds negative, but remember that Fig.~\ref{fig:Due_diligence} only shows the boundaries within which our model is accurate. This is not the same as saying that our instability does not affect larger particles. Indeed, one could argue that including vortex evolution in the model would allow $\dtg$ to increase over time, and since $\gamma$ correlates positively with $\dtg$, the instability might be stronger than our model suggests.

\subsubsection{Other limitations}
\label{ssub:Due_diligence_other_limitations}

We ignore viscosity, turbulent diffusion, collisions, anything happening at the vortex's boundary, \textit{etc.} The list of non-ideal effects is long, but this is a first investigation into dusty vortex dynamics, so it makes sense to start with a simple model.

More importantly, our analysis is limited to the onset of the instability, so we cannot say anything about its saturation, especially its ability to form dust clumps. This issue could and should be adressed using simulations, but is beyond the scope of this study.

\subsection{Comparison to simulations}
\label{sub:Discussion_comparison_to_simulations}

Several teams have run numerical experiments of dust-vortex interactions, and found that the dust's backreaction triggers an instability. One source of perplexity is that this unknown instability appears both in 2D and 3D simulations, whereas the standard \SI\ only appears in 3D. If our instability is the one from the simulations, then we can resolve that paradox. Indeed, we showed in \S\ref{sub:2D_instability} that the vortex \SI\ extends to 2D.

Unfortunately, the quantitative match between our instability and the simulations is poor. For instance, if we compare the growth rates, Fig.~\ref{fig:2D_instability} shows that 2D axisymmetric perturbations have an $\e$-folding time of 40 orbits. This is consistent with \cite{Fu+2014} who reports an instability within the first 100 orbits. But \cite{Raettig+2015}, \cite{Surville+2016} and \cite{Raettig+2021} all witness growth on the orbital timescale.

Then, consider the resolution necessary to observe the instability. Figure~\ref{fig:2D_instability} shows that the fastest growing mode of the 2D instability has a wavelength ${ \lambda / b \approx 2 \pi / 60 }$. It is generally thought that vortices can reach a size comparable to the disc's scale height~$H$ \citep{Shen+2006}. So, a simulation would need a minimum resolution of 38 cells per scale height to resolve the fastest growing mode with four cells per wavelength. More precisely, this estimate is valid when ${ \St = 10^{-2} }$. The scaling rule of \S\ref{sub:Exploration_of_parameter_space_St} implies that if $\St$ is ten times larger, then the resolution can be ten times lower. Despite this, \cite{Surville+2016}, \cite{SurvilleMayer2019} and \cite{Lovascio+2022} all have runs where they see an instability even though their resolution is well below the minimum needed to see the vortex \SI.

Those discrepancies may simply be because the simulated vortices have non-Kida pressure and vorticity profiles, leading to different dust drift speeds, growth rates, and preferred wavelengths. Alternatively, our model is only valid for small-wavelength perturbations, so maybe the instability extends to large scales but we fail to describe it. The third possibility is that the instability seen in simulations is not the vortex \SI\ but another dust-induced instability.

One piece of evidence in this direction comes from \cite{Lovascio+2022}, who report an instability only if the sound speed is low. Our model implicitly assume that the sound speed is infinite, so our instability and theirs are probably distinct.

\subsection{Applications}
\label{sub:Discussion_applications}

\subsubsection{Planet formation theory}
\label{sub:Discussion_applications_to_planet_formation_theory}

Our most important result is that the vortex instability is a form of streaming instability. As such, it most likely saturates by forming dust filaments and clumps. The clumps may then collapse gravitationally to form planetesimals, provided that they reach a high enough density. However, a non-linear simulation of the vortex \SI\ is required to validate this concept.

\subsubsection{Vortex observations}
\label{sub:Discussion_vortex_observations}

Our instability is similar to the \SI, so we expect it to clump the dust and thereby to modify the optical properties of the medium. Specifically, \cite{Scardoni+2021} found that the optically thick fraction drops as the emission from particles trapped in optically thick clumps is suppressed. \cite{Stammler+2019} further predict that if the \SI\ stops when the pebble density falls below a critical value, then the optical depth should be uniform across recently-\SI-active regions. We expect both of those predictions to remain true in vortices.

\subsubsection{Vortex evolution}
\label{sub:Discussion_vortex_evolution}

It remains unclear whether the instability destroys the vortex or not. All the 2D simulations end with the vortex's demise, but the 3D simulation of \cite{Raettig+2021} shows only the midplane vortex being destroyed while the rest of the vortex column survives. We only modelled the onset of instability, so we cannot contribute to this debate. But if the instability does destroy vortices, then our work can help estimate the `vortex life expectancy'.

\subsubsection{RDI theory}
\label{sub:Discussion_applications_to_RDI_theory}

The work of appendix~\ref{sec:Non_modal_RDI_theory} extends \RDI\ theory from sine wave to all periodic signals. Furthermore, the fact that the \SI\ remains active in vortices show that \RDIs\ are robust to complexity in the background flow and in the wave structure.

\section{Conclusion}
\label{sec:Conclusion}

This series of papers studies what happens when one adds dust to a Kida vortex embedded in a \PPD. In paper I we showed that if the vortex is weak and anticyclonic, then its centre is a pressure maximum and the dust spirals towards it. We then studied how dust concentration affects the vortex' long term (laminar) evolution. In the present paper, we showed that the dust's spiral motion can couple with the vortex’s inertial waves to trigger an instability akin to the \SI.

This instability has the potential to be an important mode of planetesimal formation. Indeed, the \SI\ is known to saturate by forming dust clumps \citep{JohansenYoudin2007}, which can collapse gravitationally to form planetesimals. Since our instability is so similar to the SI, it probably saturates in the same way. Additionally, the vortex SI may be more robust than the standard \SI. Indeed, the \SI\ can only develop in places containing a high density of similar-sized pebbles \citep{Johansen+2009, Carrera+2015, Krapp+2019}. Large-scale vortices naturally provide such an environment, thanks to their tendency to selectively trap pebbles of a certain size \citep{BargeSommeria1995}.

One key difference between the standard SI and our `vortex \SI' is that the latter remains active in 2D. We investigate whether this could explain the unknown instability seen in 2D dusty vortex simulations. Unfortunately, our instability is weak and small-scale compared to that seen in the simulations. This may be just an artifact of our vortex model, or of our focus on small-wavelength perturbations. But it could also indicate that our instability is superseded by another, faster, vortex \RDI. This hypothesis could be studied by dropping the incompressibility assumption, thereby allowing sound waves and density waves to propagate.

Because our work is analytical, we had to make many assumptions. We modelled the gas as incompressible, we assumed all the dust particles have the same size, we neglected collisions between them, we limited ourselves to the regime of dilute and small dust, we neglected viscosity and turbulent diffusion, and we neglected any effect happening at the vortex’s boundary. But arguably, the strongest limitations stem from the simplicity of our vortex model, especially our neglection of the vertical component of gravity.

Non-linear numerical simulations are needed to make progress in many of these directions. However, the small spatial scale at which the instability operates makes such simulations challenging, especially when combined with the small growth rate. This could be alleviated by running local simulations inside the `vortex shearing box’ (\S\ref{sec:VSB}).

All that being said, the analytical approach allowed us to construct a detailed physical explanation for the mechanism of the instability. Such an explanation has been lacking in all  numerical simulations to date.

\section*{Acknowledgments}

We wish to thank the anonymous referee and Andrew Youdin for their advice that helped us clarify several parts of the manuscript. Support for N.M. was provided by a Cambridge International \& Isaac Newton Studentship.

\section*{Data Availability}

The data and numerical codes underlying this article were produced
by the authors. They will be shared on reasonable request to the
corresponding author. The code is distributed on Github at the following
URL: \url{https://github.com/NathanMagnan/LAVA.git}.

\bibliographystyle{mnras}
\bibliography{main}

\appendix
\clearpage

\onecolumn
\section{The vortex shearing box}
\label{sec:VSB}

In the body of the paper, we focused on small-scale perturbations. To study them, we took inspiration from the shearing box and built a `vortex shearing box'. The present appendix describes our model.

\subsection{Geometry of the box}
\label{sub:VSB_geometry}

We parametrize the centre of the box (${ X_{0}, Y_{0} }$) as
\begin{subequations}
    \label{eq:VSB_change_of_coordinates_centre_of_box}
    \begin{align}
        X_{0} &= b \, \cos{(\Theta)} , \label{eq:VSB_change_of_coordinates_centre_of_box_x} \\
        Y_{0} &= - a \, \sin{(\Theta)} , \label{eq:VSB_change_of_coordinates_centre_of_box_y}
    \end{align}
\end{subequations} 
where $a$ and $b$ are the semi-major and semi-minor axes of the reference streamline, and ${ \Theta }$ is an angle that describes the instantaneous position of the box. \\

\noindent Since the centre of the box follows a fluid parcel along its Kida motion, we have ${ \rd_{t} \mathbf{X_{0}} = \buK }$. This leads to ${ \rd_{t} \Theta = S / (\alpha - 1) }$. Assuming without loss of generality that ${ \Theta(t = 0) = 0 }$, we get
\begin{equation}
    \label{eq:VSB_evolution}
    \Theta (t) = \frac{t S}{\alpha - 1} .
\end{equation}
\vspace{-0.1 \baselineskip}

\noindent Regarding the axes of the box, we will use ${ \ex }$ to denote the box's `radial' unit vector, and ${ \ey }$ for the `azimuthal' unit vector. We demand that ${ \ey }$ remains aligned with ${ \buK }$ and that ${ \ex }$ points away from the vortex's centre, leading to
\begin{equation}
    \label{eq:VSB_change_of_coordinates_vectors_theta}
    \ex = \frac{\alpha \cos (\Theta) \, \eX - \sin (\Theta) \, \eY}{\sqrt{\sin^{2}(\Theta) + \alpha^{2} \, \cos^{2}(\Theta)}} , \quad \quad
    \ey = - \frac{\sin (\Theta) \, \eX + \alpha \cos (\Theta) \, \eY}{\sqrt{\sin^{2}(\Theta) + \alpha^{2} \, \cos^{2}(\Theta)}} , \quad \quad
    \ez = - \eZ .
\end{equation}
Looking at the first two equations, we see that we can simplify things by introducing a new angle $\varphi$ such that \smash{${ \alpha \tan (\varphi) = \tan (\Theta) }$}:
\!\!\!\!\!\!\begin{equation}
    \label{eq:VSB_change_of_coordinates_vectors_phi}
    \ex = + \cos (\varphi) \, \eX - \sin (\varphi) \, \eY , \quad \quad
    \ey = - \sin (\varphi) \, \eX - \cos (\varphi) \, \eY , \quad \quad
    \ez = - \eZ .
\end{equation}
As explained in \S\ref{ssub:Vortex_shearing_box}, $\Theta$ tracks the position of the box and $\varphi$ tracks its orientation. All in all, the change of coordinates combines a translation, a rotation around the vertical axis, and a reflexion across the horizontal plane:
\begin{equation}
    \label{eq:VSB_change_of_coordinates_coordinates}
    x = + \cos (\varphi) (X - X_{0}) - \sin (\varphi)  (Y - Y_{0}) , \quad \quad \\
    y = - \sin (\varphi) (X - X_{0}) - \cos (\varphi) (Y - Y_{0}) , \quad \quad \\
    z = - Z .
\end{equation}
\vspace{-0.1 \baselineskip}

\noindent Note that we had a choice for the direction in which to measure $\Theta$ and $\varphi$. The current choice ensures that $\Theta$ and $\varphi$ increase over time. We also had a choice for the direction of $\ex$, $\ey$, and $\ez$. We opted to have $\ex$ directed towards the vortex's boundary -- so that $\ex$ in the vortex box has the same meaning as $\eX$ in the shearing box -- and $\ey$ in the same direction as $\buK$ -- just like~$\eY$ is in the orbital direction in the shearing box. However, since we want ${ \{ \ex, \, \ey, \, \ez \} }$ to be right-handed, we need to impose~${ \ez = - \eZ }$. This may confuse the reader. Unfortunately, there is no perfect choice for anticyclonic vortices.

\subsection{Dynamics in the box}
\label{sub:VSB_dynamics}

Using the vortex box requires a change of coordinates. Finding the new expressions of the forces that were already active in the shearing box is straightforward, we just need to translate Eqs.~\eqref{eq:Kida_vortex_in_shearing_box_P}, \eqref{eq:forces_in_shearing_box_tidal_potential} and~\eqref{eq:forces_in_shearing_box_Coriolis} via the change of variable defined by Eqs.~\eqref{eq:VSB_change_of_coordinates_centre_of_box}, \eqref{eq:VSB_change_of_coordinates_vectors_phi} and~\eqref{eq:VSB_change_of_coordinates_coordinates}. We find
\begin{subequations}
    \label{eq:old_forces_in_VSB}
    \begin{align}
        \bnabla \hK &= 
        \begin{multlined}[t]
            \big[ q_{1} X_{0} \cos(\varphi) - q_{2} Y_{0} \sin (\varphi) \big] \ex - \big[ q_{1} X_{0} \sin (\varphi) + q_{2} Y_{0} \cos (\varphi) \big] \ey \\
            \hspace{2 cm} + \big[ q_{1} \cos^{2} (\varphi) + q_{2} \sin^{2} (\varphi) \big] x \ex + \big[ q_{1} \sin^{2} (\varphi) + q_{2} \cos^{2} (\varphi) \big] y \ey + \big[ q_{2} - q_{1} \big] \frac{\sin (2 \varphi)}{2} \left( x \ey + y \ex \right) , 
        \end{multlined} \!\!\!\! \label{eq:forces_in_VSB_P} \\
        - \bnabla \Phi_{t} &= 2 \Omega S \, \bigg\{ X_{0} \left[ \cos (\varphi) \ex - \sin (\varphi) \ey \right] + \cos^{2} (\varphi) \, x \ex + \sin^{2} (\varphi) \, y \ey - \frac{\sin (2 \varphi)}{2} \left( x \ey + y \ex \right) \bigg\} , \label{eq:forces_in_VSB_tidal_potential} \\
        \mathbf{f_{Co}^{\text{s}}} &= 2 \Omega \, \ez \wedge \bu_{g \text{ or } d} . \label{eq:forces_in_VSB_Coriolis_from_SB}
    \end{align}
\end{subequations}
\vspace{-0.1 \baselineskip}

\newcommand{\bOmegavs}{\bOmega_{\text{v} / \text{s}}} 

\enlargethispage{-1 \baselineskip}
\noindent Formally, the Coriolis force from the shearing box ${ \mathbf{f_{Co}^{\text{s}}} }$ contains a second term, but we prefer to set it apart because it depends on position rather than speed. We call it the \textit{composition term} because it emerges when one composes two changes of reference frame. Its expression is
\begin{equation}
    \label{eq:forces_in_VSB_composition_force}
    \mathbf{f_{co}^{\text{v-s}}} = - 2 \Omega \, \eZ \wedge \left( \rd_{t} \mathbf{X_{0}} + \bOmegavs \wedge \bx \right) = - \frac{2 \Omega S}{\alpha - 1} \sqrt{\left(\alpha X_{0}\right)^{2} + \left(Y_{0} / \alpha\right)^{2}} \, \ex - 2 \Omega \, (\rd_{t} \varphi) \left( x \ex + y \ey \right) .
\end{equation}
where ${ \bOmegavs = \rd_{t} \varphi \, \ez = - \rd_{t} \varphi \, \eZ }$ is the rotation vector of the vortex box relative to the shearing box. Interestingly, one can show that \smash{${ \rd_{t} \varphi = \frac{S}{\alpha - 1} \left[ \alpha \sin^{2} (\varphi) + \cos^{2} (\varphi) / \alpha \right] }$}. This will help simplify Eqs.~\eqref{eq:new_forces_in_VSB}. \\

\noindent As explained in \S\ref{sub:Linear_perturbation_equations_small_wavelength_regime}, since the vortex box is accelerating with respect to the shearing box, we need to account for a new array of fictitious forces: a second centrifugal force ${ \mathbf{f_{ce}^{\text{v}}} }$ (we attach the translational fictitious force ${ \rd_{t}^{2} \mathbf{X_{0}} }$ for convenience), a second Coriolis force ${ \mathbf{f_{Co}^{\text{v}}} }$, and an Euler force ${ \mathbf{f_{Eu}^{\text{v}}} }$. We find the following expressions
\begin{subequations}
    \label{eq:new_forces_in_VSB}
    \begin{align}
        & \mathbf{f_{ce}^{\text{v}}} \! = \! - \rd_{t}^{2} \mathbf{X_{0}} \! - \! \bOmegavs \! \wedge \! (\bOmegavs \! \wedge \! \bx) \! = \! \left( \frac{S}{\alpha \! - \! 1} \right)^{2} \bigg\{ \left[X_{0} \cos (\varphi) \! - \! Y_{0} \sin (\varphi)\right] \ex \! - \! \left[X_{0} \sin (\varphi) \! + \! Y_{0} \cos (\varphi)\right] \ey \bigg\} \! + \! (\rd_{t} \varphi)^{2} \left( x \ex \! + \! y \ey \right) \! , \label{eq:forces_in_VSB_centrifugal_from_VSB} \\
        & \mathbf{f_{Co}^{\text{v}}} = -2 \, \bOmegavs \wedge \bu_{g \text{ or } d} = - 2 \, (\rd_{t} \varphi) \, \ez \wedge \bu_{g \text{ or } d} , \label{eq:forces_in_VSB_Coriolis_from_VSB} \\
        & \mathbf{f_{Eu}^{\text{v}}} = - (\rd_{t} \bOmegavs) \wedge \bx = (\rd_{t}^{2} \varphi) \left( y \, \ex - x \, \ey \right) . \label{eq:forces_in_VSB_Euler} 
    \end{align}
\end{subequations}
\vspace{-0.1 \baselineskip}

\noindent Finally, the Kida flow~\eqref{eq:Kida_vortex_in_shearing_box} becomes ${ \buK^{\text{v}} = \buK^{\text{s}} - (\rd_{t} \mathbf{X_{0}} + \mathbf{\Omega}_{\text{v} / \text{s}} \wedge \bx) }$, leading to formula~\eqref{eq:Kida_vortex_in_VSB_u}.

\vspace{-0.5 \baselineskip}
\subsection{Simplification of the linear perturbation equations in the small-wavelength regime}
\label{sub:VSB_simplifications}

In \S\ref{ssub:Linear_perturbation_equations_small_wavelength_regime_simplifications}, we claimed that using the vortex shearing box and assuming small-wavelength perturbations, we could simplify the
linear perturbation equations~\eqref{eq:linear_perturbation_equations}. This subsection presents our method.

\newcommand{\bvou}{\underline{\bvo}} 

Firstly, we decompose the background dust-to-gas drift ${ \bvo = \tau \, \bnabla \hK }$ into a uniform part \smash{${ \bvob }$} and a space-dependent part \smash{${ \bvou (\bx) }$}. This decomposition is useful because $\bvob$ scales with $b$, whereas $\bvou$ scales with $x$, so if the box is small compared to the semi-minor axis of the streamline it is attached to, we always have ${ x \ll b }$ and we can neglect $\bvou$.

\newcommand{\buKb}{\overline{\buK}}
\newcommand{\buKu}{\underline{\buK}}
\newcommand{\gradhKb}{\overline{\bnabla \hK}}
\newcommand{\gradhKu}{\underline{\bnabla \hK}}

\newcommand{\bugordi}{\bu_{g \text{ or } d, 1}}

\newcommand{\tbuKu}{\underline{\check{\bu}_{K}}}
\newcommand{\tbvob}{\check{\overline{\bv}}_{0}}
\newcommand{\tbvou}{\underline{\check{\bv}_{0}}}

\renewcommand{\tbugi}{\check{\bu}_{g, 1}}
\renewcommand{\tbudi}{\check{\bu}_{d, 1}}
\newcommand{\tbugordi}{\check{\bu}_{g \text{ or } g, 1}}
\newcommand{\tbvi}{\check{\bv}_{1}}

\newcommand{\tbnabla}{\check{\bnabla}}

The next step is to adimensionalise the variables. We do so according to Eqs.~\eqref{eq:Kida_vortex_in_VSB_u} and~\eqref{eq:forces_in_VSB_P},
\begin{equation}
    \label{eq:vortex_simplification_at_small_scales_adimensional_variables}
    \begin{aligned}
        \buKb &= 0, && & \buKu &= \Omega \, x \, \tbuKu, && & \bugordi &= \Omega \, b \, \tbugordi, \\
        \gradhKb &= \Omega^{2} \, b \, \tbvob, \quad   && & \gradhKu &= \Omega^{2} \, x \, \tbvou, \quad&& & \bnabla \hi &= \Omega^2 \, b \, \tbvi, \\
        \bnabla \overline{f} &= 0 , && & \bnabla \underline{f}  &= x^{-1} \, \underline{\check{f}}, && & \bnabla f_{1} &= \lambda^{-1} \, \check{f}_{1} ,
    \end{aligned}
\end{equation}
where $f$ represents a dimensional variable and ${\check{f}}$ its adimensional counterpart. With these notations, equation~\eqref{eq:linear_pertubed_continuity_dust} becomes
\begin{equation}
    \nonumber
    \partial_{t} \rdi + \Omega \, \rdi \tbnabla \bcdot \tbuKu + \St \, \Omega \, \rdi \tbnabla \bcdot \tbvou + \Omega \, \frac{x}{\lambda} \, \tbuKu \bcdot \tbnabla \rdi + \St \, \Omega \, \frac{b}{\lambda} \, \tbvob \bcdot \tbnabla \rdi + \St \, \Omega \, \frac{x}{\lambda} \, \tbvou \bcdot \tbnabla \rdi = - \Omega \, \frac{a}{\lambda} \, \tbnabla \bcdot \tbudi .
\end{equation}
Firstly, the Kida flow is incompressible, so the second term is null. Secondly, we can neglect the sixth term with respect to the fifth if ${ x \ll b }$. Finally, we can neglect the third term with respect to the fifth if ${ \lambda \ll b }$. This outlines a set of conditions under which Eq.~\eqref{eq:linear_pertubed_continuity_dust} rigorously simplifies to
\begin{equation}
    \label{eq:simplification_at_small_scales_example}
    \partial_{t} \rdi + (\buK + \tau \, \gradhKb)  \bcdot \bnabla \rdi = - \bnabla \bcdot \budi .
\end{equation}
Following this example, Eqs.~\eqref{eq:linear_perturbation_equations} simplify to Eqs.~\eqref{eq:linear_pertubation_equations_small_wavelength}.

\enlargethispage{+ 3 \baselineskip}
\thispagestyle{empty}
At this stage, we can justify the statement made in \S\ref{sub:Linear_perturbation_equations_small_wavelength_regime} that the space-dependence of ${ \bugo }$ and ${ \budo }$ precludes Fourier decomposition. What we mean is that we would like to use the trick from Eq.~\eqref{eq:wavevector_evolution} to make all coefficients in the Fourier-transformed equations independent of $\bx$. This trick is essentially a choice of coordinate system that accounts for the shear. Such a coordinate system exists if the background flow is linear in space. Now ${ \bugo }$ and ${ \bugo }$ are both linear, so one could think that the trick works. Unfortunately, the two flows are different, so no choice of coordinates could account for both shears at the same time. What we do in Eq.~\eqref{eq:simplification_at_small_scales_example} is show that on small scales, the difference between the two flows is dominated by the drift, so we can neglect the difference in shear. This is what allows us to use the trick from Eq.~\eqref{eq:wavevector_evolution}.

\begin{figure}
    \centering

    \begin{minipage}{0.45 \textwidth}
        \centering
        \includegraphics[width = \linewidth]{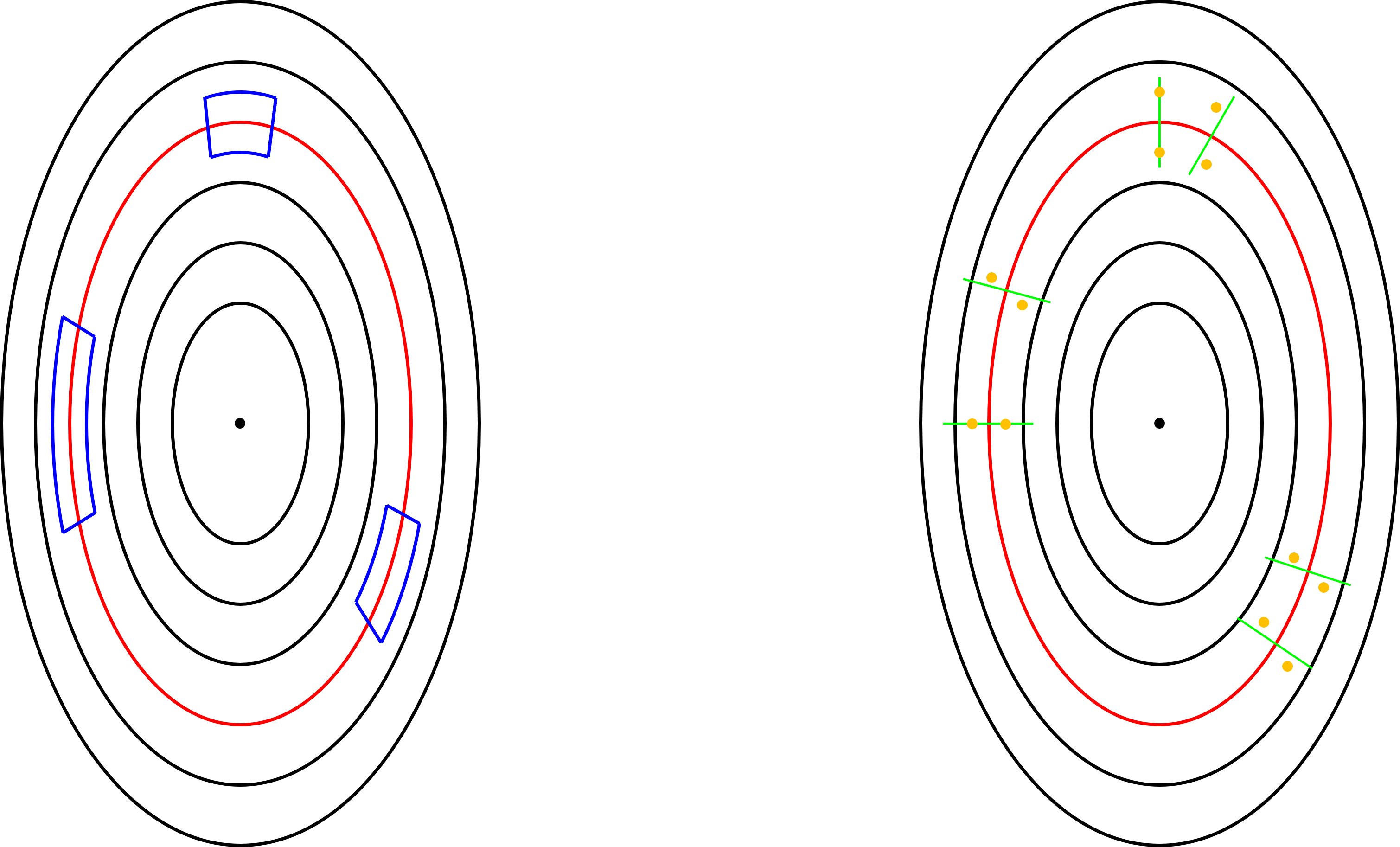}
        \caption{Diagrams explaining the origin of strain and shear. In black are the streamlines of a Kida vortex. The vortex shearing box accompanies a reference fluid parcel as it follows a particular streamline, in red. 
        \textit{Left:} The blue boxes represent the deformation of an incompressible fluid parcel due to the non-uniformity of elliptical motion. This is the source of strain. 
        \textit{Right:} The orange dots represent two fluid parcels at different times, and the green lines represent the `radial' axis of the vortex box at those times. This explains the~shear.}
        \label{fig:Shear_and_Strain}
    \end{minipage}
    \hfill
    \begin{minipage}{0.52 \textwidth}
        \centering
        \vspace{0.4 cm}
        \includegraphics[width = \linewidth]{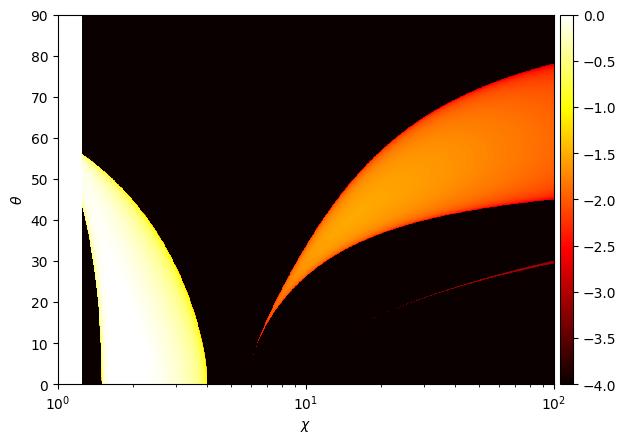}
        \caption{
            Figure~2 from \protect\cite{LesurPapaloizou2009}, reproduced with \nameofmycode. It displays the growth rate of the \EI\ versus the wavevector's latitude and the vortex's aspect ratio.
        }
        \label{fig:Lesur_Papaloizou}
    \end{minipage}

\end{figure}

\twocolumn
\section{The numerical solver}
\label{sec:LAVA}

As stated earlier, the simplified perturbation equations~\eqref{eq:linear_perturbation_equations_Fourier_transformed} are still too complex to be solved analytically (at least in 3D). So, we need to develop and implement an algorithm to solve these equations. We call our code \nameofmycode\ for \textit{Linear Analysis of Vortices in Astrophysical discs}. The goal of the present appendix is to explain how it works.

In \S\ref{sub:LAVA_nature_of_the_equations} we analyse the structure of Eqs.~\eqref{eq:linear_perturbation_equations_Fourier_transformed}, and provide methods to overcome the structural difficulties. Then in \S\ref{sub:LAVA_algorithm} we explain how \nameofmycode\ combines these methods
to compute the growth rate of the instability. Finally, in \S\ref{sub:LAVA_validation} we run two tests to make sure that \nameofmycode\ works as intended.

\vspace{-1 \baselineskip}
\subsection{Nature of the equations to solve}
\label{sub:LAVA_nature_of_the_equations}

\subsubsection{A Floquet problem}
\label{ssub:LAVA_nature_of_the_equations_Floquet_problem}

Most of the coefficients in Eqs.~\eqref{eq:linear_perturbation_equations_Fourier_transformed} involve $\bk$, $\bvob$, $\bA$ or $\bOmega$, all of which are time-dependent. This is because the vortex is elliptic, so the background flow inside the `vortex box' is not the same at vertex and co-vertex. Fortunately, since the vortex's streamlines are closed, those coefficients are periodic.

This makes Eqs.~\eqref{eq:linear_perturbation_equations_Fourier_transformed} a Floquet problem. To determine the Floquet exponents, one can choose a linear basis of initial values ${ \{ \mathbf{y_{1}^{i.}}, \mathbf{y_{2}^{i.}}, ..., \} }$, solve the \IVP\ for each initial value, and consider the final values after one period ${ \{ \mathbf{y_{1}^{f.}}, \mathbf{y_{2}^{f.}}, ..., \} }$. Decomposing the final values over the basis of initial values provides the monodromy matrix of the Floquet problem, ${ \mathbf{R}(2 \pi) }$. The eigenvalues of this matrix are the characteristic multipliers ${ \e^{2 \pi \gamma} }$.

\vspace{-1 \baselineskip}
\subsubsection{A differential-algebraic equation}
\label{ssub:LAVA_nature_of_the_equations_DAE}

Note also that Eq.~\eqref{eq:Fourier_transformed_equations_continuity_gas} is algebraic, whereas Eqs.~(\ref{eq:Fourier_transformed_equations_continuity_dust}, \ref{eq:Fourier_transformed_equations_momentum_gas}, \ref{eq:Fourier_transformed_equations_momentum_dust}) are differential. This qualifies Eqs.~\eqref{eq:linear_perturbation_equations_Fourier_transformed} as a \DAE. Problems of this kind are common in incompressible fluid dynamics.

\newcommand{\by}{\mathbf{y}} 

\newcommand{\byd}{\mathbf{y_{d}}} 
\newcommand{\byc}{\mathbf{y_{c}}} 

\newcommand{\bHd}{\mathbf{H_{d}}} 
\newcommand{\bHc}{\mathbf{H_{c}}} 
\newcommand{\bCd}{\mathbf{C_{d}}} 

We decompose ${ \by  = \{ \thi, \tbugi, \trdi, \tbudi \} }$ into the dynamical variables ${ \byd = \{ \tbugi, \trdi, \tbudi,  \} }$ and the constrained variable ${ \byc = \thi }$. Equations~\eqref{eq:linear_perturbation_equations_Fourier_transformed} can be rewritten as
\begin{subequations}
    \label{eq:DAE_index_2}
    \begin{align}
        \rd_{t} \byd &= \bHd \, \byd + \bHc \, \byc , \label{eq:DAE_index_2_1} \\
        \mathbf{0} &= \, \bCd \, \byd . \label{eq:DAE_index_2_2}
    \end{align}
\end{subequations}
Differentiating Eq.~\eqref{eq:DAE_index_2_2} yields ${ \bCd \, (\rd_{t} \byd) + (\rd_{t} \bCd) \, \byd = 0 }$. Injecting Eq.~\eqref{eq:DAE_index_2_1} then shows that ${ \byc }$ is solution to
\begin{equation}
    \bCd \, \bHc \, \byc = - (\rd_{t} \bCd + \bCd \, \bHd) \, \byd , \label{eq:DAE_linear_problem}
\end{equation}
which is a simple linear problem. Therefore, we can use an off-the-shelf \ODE\ solver on ${ \byd }$, but each time it calls for the derivative of ${ \byd }$, we start by solving Eq.~\eqref{eq:DAE_linear_problem}. This provides the correct value of ${ \byc }$ to input into Eq.~\eqref{eq:DAE_index_2_1}, which finally gives the desired derivative. 

This method can be stiff, so we use an \ODE\ solver based on the backward differentiation formula of \cite{CurtissHirschfelder52}. Specifically, we use the implementation offered by Scipy's \verb|solve_ivp| routine \citep{Scipy, ShampineReichelt97}

\vspace{-1 \baselineskip}
\subsubsection{A differential-algebraic Floquet problem}
\label{ssub:LAVA_nature_of_the_equations_Floquet_DAE}

The method described in \S\ref{ssub:LAVA_nature_of_the_equations_DAE} only conserves the value of ${ \bCd \, \byd }$, so Eq.~\eqref{eq:DAE_index_2_2} is verified only if it was verified by the initial value. Consequently, we need a basis of \textit{acceptable} initial values for the Floquet analysis of \S\ref{ssub:LAVA_nature_of_the_equations_Floquet_problem}.

Remember that Eq.~\eqref{eq:DAE_index_2_2} represents Eq.~\eqref{eq:Fourier_transformed_equations_continuity_gas}, which is really scalar rather than vectorial. Therefore, we can rewrite Eq.~\eqref{eq:DAE_index_2_2} as the dot product ${ \hat{\mathbf{e}} \bcdot \byd = 0 }$, and use Schmidt's procedure to complete ${ \hat{\mathbf{e}} }$ into an orthonormal basis of ${ \mathbb{R}^{7} }$. The last 6 members of this basis are the 6 desired \textit{acceptable} initial values ${ \{ \mathbf{y_{1}^{i.}}, ..., \mathbf{y_{6}^{i.}}\} }$.

\vspace{-1 \baselineskip}
\subsection{\nameofmycode's algorithm}
\label{sub:LAVA_algorithm}

First, \nameofmycode\ uses Schmidt's procedure to obtain an orthonormal basis of acceptable initial values ${ \{ \mathbf{y_{1}^{i.}}, ..., \mathbf{y_{6}^{i.}}\} }$. Then, it evolves these initial values for one period, using \verb|solve_ivp| and the index-reduction method described in \S\ref{ssub:LAVA_nature_of_the_equations_DAE}. This way, the code finds the 6 final values ${ \{ \mathbf{y_{1}^{f.}}, ..., \mathbf{y_{6}^{f.}} \} }$. Since the initial basis was orthonormal, it is straightforward to decompose the final values and obtain the monodromy matrix ${ \mathbf{R}(2 \pi)_{m, n} = \mathbf{y_{m}^{i.}} \bcdot \mathbf{y_{n}^{f.}} }$. Finally, \nameofmycode\ computes the eigenvalues of this matrix, deduces the Floquet exponents ${ \{ \gamma_{1}, ..., \gamma_{6} \} }$, and returns the growth rate ${ \mathrm{Re}(\gamma) }$ of the fastest growing mode. This is the growth rate of the instability.

\begin{figure}
    \centering
    \includegraphics[width = \linewidth]{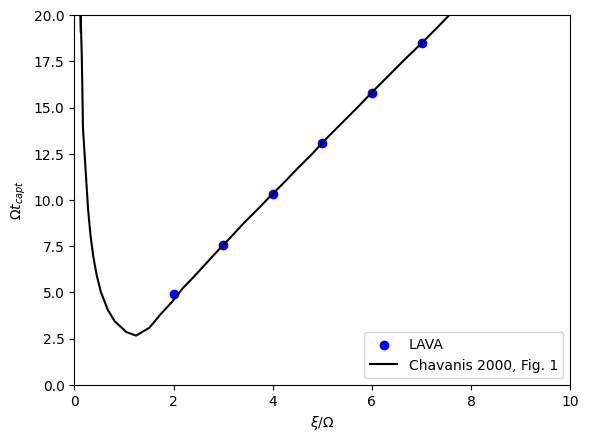}
    \caption{
        Figure~1 from \protect\cite{Chavanis2000}, partially reproduced with \nameofmycode. It displays the e-folding time of a particle's inspiral towards the vortex's centre, as a function of that particle's inverse Stokes number.
    }
    \label{fig:Chavanis}
    \vspace{-1 \baselineskip}
\end{figure}

\vspace{-1 \baselineskip}
\subsection{Code validation}
\label{sub:LAVA_validation}

\subsubsection{Reproduction of the elliptical instability}
\label{ssub:LAVA_validation_EI}

To test \nameofmycode, we try and reproduce Figure~2 from \cite{LesurPapaloizou2009}. Reproducing the \EI\ is a good test for \nameofmycode\, because it means solving Eqs.~\eqref{eq:linear_perturbation_equations_Fourier_transformed} without dust. We obtain exactly the expected results, as shown in Fig.~\ref{fig:Lesur_Papaloizou}. This validates \nameofmycode's algorithm and implementation thereof.

\vspace{-1 \baselineskip}
\subsubsection{Reproduction of the dust concentration rate}
\label{ssub:LAVA_validation_dust_concentration}

We also try and reproduce Figure~1 from \cite{Chavanis2000}. This is a good test because it allows us to verify that the dust part of the equations of motion is correct.

We could only perform the test for small particles, because marginally-coupled and large particles follow complex trajectories whose instantaneous semi-minor axes are hard to estimate. Nevertheless, for ${ \St \ll 1 }$ particles, we obtain exactly the expected results, as shown in Fig.~\ref{fig:Chavanis}. This confirms that we did not forget or miscalculate any force when deriving the `vortex shearing box'.

\section{The response of test particles to zonal flows}
\label{sec:Response_of_test_particles_to_zonal_flows}

In \S\ref{ssub:2D_instability_why}, we proposed an approximation for the velocity of test particles embedded in a (vortex-wise) zonal flow. The goal of the present appendix is to justify this approximation.

\newcommand{\bI}{\mathbf{I}} 

\renewcommand{\tbugi}{\tilde{\bu}_{g, 1}} 
\renewcommand{\tbudi}{\tilde{\bu}_{d, 1}} 

Firstly, in the regime of test particles, the dust equations~(\ref{eq:Fourier_transformed_equations_continuity_dust}, \ref{eq:Fourier_transformed_equations_momentum_dust}) become
\begin{subequations}
    \label{eq:linear_perturbation_equations_Fourier_transformed_dust_forced_by_inertial_wave}
    \begin{align}
        & \rd_{t} \trdi + i \, ( \bk \bcdot \bvob ) \, \trdi = - i \bk \bcdot \tbudi , \label{eq:Fourier_transformed_equations_dust_forced_by_inertial_wave_continuity} \\ 
        & \rd_{t} \tbudi + \left[ \frac{\bI}{\tau} + i \, ( \bk \bcdot \bvob ) \, \bI + \bA  + 2 \bOmega \right] \tbudi = \frac{\tbugi}{\tau} , \label{eq:Fourier_transformed_equations_dust_forced_by_inertial_wave_momentum}
    \end{align}
\end{subequations}
where $\bI$ is the identity matrix. Note the structure of those equations: $\trdi$ is controlled by $\tbudi$, which is in turn controlled by $\tbugi$. Therefore, we can start by solving Eq.~\eqref{eq:Fourier_transformed_equations_dust_forced_by_inertial_wave_momentum}.

\newcommand{\tbudip}{\tbudi'} 

In the regime of small particles, the term in ${ \bI / \tau }$ dominates the dust's response and quickly damps discrepancies between $\tbudi$ and $\tbugi$. To address the leading-order pinning of $\tbudi$ to $\tbugi$, we change the variable from $\tbudi$ to ${ \tbudip = \e^{t/\tau} \, \tbudi }$. The new variable solves
\begin{equation}
    \rd_{t} \tbudip + \left[ i \, ( \bk \bcdot \bvob ) \, \bI + \bA  + 2 \bOmega \right] \tbudip = \frac{\e^{t/\tau}}{\tau} \, \tbugi . \nonumber
\end{equation}

\newcommand{\bM}{\mathbf{M}} 
\newcommand{\fgwidsjf}{\mathbf{f}} 
\newcommand{\bU}{\mathbf{U}} 

This is a linear forced \ODE\ whose response matrix \smash{${ \bM = i \, ( \bk \bcdot \bvob ) \, \bI \! + \! \bA  \! + 2 \! \bOmega }$} and forcing vector \smash{${ \fgwidsjf = \tau^{-1} \, \e^{t/\tau} \, \tbugi }$} are time-dependent. The general form of the solution is
\begin{equation}
    \tbudip (t) = \bU (t) \, \left[ \tbudip (0) + \int_{0}^{t} \rd s \,\, \bU^{-1} (s) \, \fgwidsjf (s) \right] . \nonumber
\end{equation}
where $\bU$ is the fundamental matrix of the homogeneous equation. $\bU$ and $\tbugi$ are smooth functions of $t$, but the exponential $\e^{t/\tau}$ explodes. We can use this separation of timescales to approximate the integral. Specifically, we can apply the following lemma:

\noindent \rule{\linewidth}{0.4pt}

\noindent Let ${ f \in \mathcal{C}^{n+1} (\mathbb{R}^{+}, \mathbb{R}) }$ and ${ M > 0 }$ such that ${ \forall k \! \in \! \llbracket 0, n \! + \! 1 \rrbracket}$, ${ || f^{(k)} || \leq M }$. Then 
\begin{equation}
    \int_{0}^{t} \rd s \,\, f (s) \, \e^{s / \epsilon} \!\!\! \stackunder[5pt]{= \,\,\,\,}{$\epsilon \rightarrow 0^{+}$} \!\!\! \epsilon \, \e^{t / \epsilon} \times \sum_{k = 0}^{n} (- 1)^{k} \epsilon^{k} f^{(k)} (t) \, + \, \mathcal{O}(\epsilon^{n+2} \, \e^{t/\tau}) , \nonumber
\end{equation}
and this convergence is uniform over ${ t \in \mathbb{R}^{+} }$.

\noindent \rule{\linewidth}{0.4pt}

\noindent This leads to
\begin{multline}
    \tbudi (t) \approx \e^{-t/\tau} \, \bU (t) \, \tbugi (0) \\
    + \sum_{k = 0}^{n} (-1)^{k} \, \tau^{k} \, \bU(t) \left[ s \mapsto \bU^{-1} (s) \, \tbugi (s) \right]_{|t}^{(k)} , \hspace{-0.3cm} \nonumber
\end{multline}
where the term between square brackets reads as ``the function which maps $s$ to ${ \bU^{-1} (s) \, \tbugi (s) }$ is differentiated $k$ times, then evaluated in ${ s = t }$.'' The first term is quickly damped, so the previous expression simplifies to
\begin{equation}
    \label{eq:TVA_in_vortices_any_order}
    \tbudi (t) \approx \sum_{k = 0}^{n} (-1)^{k} \, \tau^{k} \, \bU(t) \left[ s \mapsto \bU^{-1} (s) \, \tbugi (s) \right]_{|t}^{(k)} .
\end{equation}
Since ${ \rd_{t} \bU^{-1} \! = \! - \bU^{-1} \, \bU^{(1)} \, \bU^{-1} }$ and ${ \bU^{(1)} \! = \! - \bM \bU }$, the ${ n = 1 }$ expansion reduces to
\begin{equation}
    \label{eq:TVA_in_vortices}
    \tbudi = \tbugi - \tau \left[ \bM \, \tbugi + \rd_{t} \tbugi \right] + \mathcal{O}(\St^{2}) .
\end{equation}
This resembles the terminal velocity approximation. Indeed, $\bM$ represents the forces competing with drag. The difference is that the vortex's periodicity introduces a history term.


This approximation is valid for any gas perturbation $\tbugi$, but in \S\ref{ssub:2D_instability_why} we only care about the effect of zonal flows. Injecting Eq.~\eqref{eq:zonal_flow} into Eq.~\eqref{eq:TVA_in_vortices} gives Eq.~\eqref{eq:dust_forced_by_zonal_flow_velocity}, as expected.

\section{Non-modal RDI theory}
\label{sec:Non_modal_RDI_theory}

The waves that propagate in the vortex are not simple sine functions of time, but we saw in \S\ref{sec:Instability} that they can still interact to drive an \RDI. To build our case, we relied on a generalisation of \RDI\ theory to non-modal waves. The goal of the present appendix is to introduce that theory. However, we shall not cover the general case, because it would require a heavy formalism that would impede clarity. We prefer to present our theory through an example. Specifically, we shall focus on the 2D axisymmetric instability of~\S\ref{sub:2D_instability}.

\newcommand{\vobx}{\overline{v}_{0, x}} 

\newcommand{\tugix}{\tilde{u}_{g, 1, x}} 
\newcommand{\tudiy}{\tilde{u}_{d, 1, y}} 

\renewcommand{\tbugi}{\tilde{\bu}_{g, 1}} 
\renewcommand{\tbudi}{\tilde{\bu}_{d, 1}} 

\newcommand{\hbf}{\hat{\mathbf{f}}} 

\newcommand{\omegaO}{\omega_{0}} 
\newcommand{\omegaI}{\omega_{1/2}} 
\newcommand{\omegaII}{\omega_{1}} 

\newcommand{\gammaO}{\gamma_{0}} 
\newcommand{\gammaI}{\gamma_{1/2}} 
\newcommand{\gammaII}{\gamma_{1}} 

\newcommand{\hbfO}{\hbf_{0}} 
\newcommand{\hbfI}{\hbf_{1/2}} 
\newcommand{\hbfII}{\hbf_{1}} 

\newcommand{\hhO}{\hat{h}_{0}} 
\newcommand{\hhI}{\hat{h}_{1/2}} 
\newcommand{\hhII}{\hat{h}_{1}} 

\newcommand{\hbugO}{\hat{\bu}_{g, 0}} 
\newcommand{\hbugI}{\hat{\bu}_{g, 1/2}} 
\newcommand{\hbugII}{\hat{\bu}_{g, 1}} 

\newcommand{\hrdO}{\hat{\varrho}_{d, 0}} 
\newcommand{\hrdI}{\hat{\varrho}_{d, 1/2}} 
\newcommand{\hrdII}{\hat{\varrho}_{d, 1}} 

\newcommand{\hbudO}{\hat{\bu}_{d, 0}} 
\newcommand{\hbudI}{\hat{\bu}_{d, 1/2}} 
\newcommand{\hbudII}{\hat{\bu}_{d, 1}} 

In order to solve the eigenvalue problem posed by Eqs.~\eqref{eq:linear_perturbation_equations_Fourier_transformed}, we need to select an ansatz for the solution. Floquet's theorem affirms that the most unstable perturbation is of the form ${ \tilde{\mathbf{f}}_{1} (t) = \hbf (t) \, \e^{(\gamma + i \omega) t} }$ where $\hbf$ is $T_v$-periodic and ${ \omega, \, \gamma \in \mathbb{R} }$. Standard \RDI\ theory suggests an expansion in powers of $\sqrt{\dtg}$,
\!\!\begin{equation}
    \label{eq:Ansatz}
    \begin{aligned}
        \omega  &= \omegaO  &&+ \sqrt{\dtg} \, \omegaI  &&+ \dtg \, \omegaII &&+ ... \, , \\
        \gamma  &= \gammaO  &&+ \sqrt{\dtg} \, \gammaI  &&+ \dtg \, \gammaII &&+ ... \, , \\
        \hbf    &= \, \hbfO &&+ \sqrt{\dtg} \, \hbfI    &&+ \dtg \, \hbfII &&+ ... \, ,
    \end{aligned}
\end{equation}
where ${ \hhO, \, \hbugO \text{ and } \hbudO }$ are all equal to zero but not $\omegaO$ nor~$\hrdO$. Finally, we assume that the instability grows slowly, in the sense that ${ \gammaO = 0 }$.

\subsection{Order \texorpdfstring{$\dtg^{0}$}{} in the dust equations}
\label{sub:Non_modal_RDI_theory_order_0_dust}

\newcommand{\trdO}{\tilde{\varrho}_{d, 0}} 

At order $\dtg^{0}$, the dust continuity equation~\eqref{eq:Fourier_transformed_equations_continuity_dust} becomes
\begin{equation}
    \rd_{t} \trdO + i \, ( \bk \bcdot \bvob ) \, \trdO = 0 , \nonumber
\end{equation}
where ${ \trdO = \hrdO \, \e^{i \omegaO t} }$ is a convenient notation. This equa-

\noindent tion is identical to Eq.~\eqref{eq:dust_density_wave_equation}, so we can jump straight to the solution:
\begin{subequations}
    \label{eq:2D_axisymmetric_instability_dust_order_0_solution}
    \begin{align}
        \omegaO &= \omega_{\text{ddw}} + \frac{2 \, n_{1} \, \pi}{T_v} , \label{eq:2D_axisymmetric_instability_dust_order_0_frequency} \\
        \hrdO   &= \tRd \, F_{\text{ddw}} (t) , \label{eq:2D_axisymmetric_instability_dust_order_0_density}
    \end{align}
\end{subequations}
This means that our ansatz, in which the perturbation in dust density dominates, isolates the dust density wave. The expansion in powers of $\sqrt{\dtg}$ is a way to study what this wave becomes outside the regime of test particles.

\subsection{Order \texorpdfstring{$\sqrt{\dtg}$}{} in the gas equations}
\label{sub:Non_modal_RDI_theory_order_half_gas}

\newcommand{\tbugI}{\tilde{\bu}_{g, 1/2}} 
\newcommand{\thI}{\tilde{h}_{1/2}} 

At order $\sqrt{\dtg}$, the gas equations~(\ref{eq:Fourier_transformed_equations_continuity_gas},~\ref{eq:Fourier_transformed_equations_momentum_gas}) become
\begin{align}
    i \bk \bcdot \tbugI &= \,\, 0 , \nonumber \\
    \rd_{t} \tbugI + \bA \, \tbugI &= - i \bk \, \thI - 2 \bOmega \, \tbugI . \nonumber
\end{align}
where ${ \tbugI = \hbugI \, \e^{i \omegaO t} }$ and ${ \thI = \hhI \, \e^{i \omegaO t} }$ are convenient notations. Those equations are identical to Eqs.~\eqref{eq:Fourier_transformed_equations_gas_only}, so the leading-order gas perturbation is a zonal flow,
\begin{subequations}
    \label{eq:2D_axisymmetric_instability_gas_order_half_solution}
    \begin{align}
        \omegaO &= 0 + \frac{2 \, n_{2} \, \pi}{T_v} , \label{eq:2D_axisymmetric_instability_gas_order_half_frequency} \\
        \hbugI &= \tUgy \, F_{\text{zf}} \, \ey . \label{eq:2D_axisymmetric_instability_gas_order_half_velocity}
    \end{align}
\end{subequations}
Note that Eq.~\eqref{eq:2D_axisymmetric_instability_dust_order_0_frequency} and Eq.~\eqref{eq:2D_axisymmetric_instability_gas_order_half_frequency} only agree when 
\begin{equation}
    \label{eq:resonance_condition_2D}
    \exists \, n \in \mathbb{Z} \text{ such that } \omega_{\text{ddw}} = \frac{2 n \pi}{T_v} . 
\end{equation}
Since ${ \omega_\text{iw} = 0 }$, this is equivalent to Eq.~\eqref{eq:resonance_condition}. It generalises the \RDI\ resonance condition to non-modal waves. It also means that our ansatz in powers of $\sqrt{\dtg}$ is only valid at resonance.

\subsection{Order \texorpdfstring{$\sqrt{\dtg}$}{} in the dust equations}
\label{sub:Non_modal_RDI_theory_order_half_dust}

At order $\sqrt{\dtg}$, the dust equations become
\begin{align}
    & (\gammaI + i \omegaI) \, \hrdO + \rd_{t} \hrdI + i \, ( \bk \bcdot \bvob ) \, \hrdI = - i \bk \bcdot \hbudI , \nonumber \\ 
    & \rd_{t} \hbudI + \! \left[ \frac{\bI}{\tau} \! + \! i \, ( \bk \bcdot \bvob ) \, \bI \! + \! \bA  \! + \! 2 \bOmega \right] \hbudI = \frac{\hbugI}{\tau} , \nonumber
\end{align}
Note that we do not need to introduce Doppler-shifted terms like \smash{$\trdO$} or \smash{$\tbugI$} anymore, because we now know that ${\omegaO = 0}$. 

The momentum equation is identical to Eq.~\eqref{eq:Fourier_transformed_equations_dust_forced_by_inertial_wave_momentum}, so we can expect the next-order dust velocity perturbation to be that of a dusty zonal flow:
\begin{equation}
    \hbudI = 2 \tau \left[ \left(\rd_{t} \varphi\right) - \Omega \right] \tUgy \, F_{\text{zf}} = \tUgy \, F_{\text{dzf}} . \nonumber
\end{equation}
Injecting this into the continuity equation gives
\begin{multline}
    \nonumber
    (\gammaI + i \omegaI) \, \tRd \, F_{\text{ddw}} + \rd_{t} \hrdI + i \, ( \bk \bcdot \bvob ) \, \hrdI \\ = - i \, \tUgy \, k_{x} \, F_{\text{dzf}} . \hspace{-0.3cm}
\end{multline}

To eliminate $\hrdI$, one strategy is to take the Hermitian product of the equation with a function that is orthogonal to the image of the operator ${ \rd_{t} + i \, (\bk \bcdot \bvob) \, \bI }$. $F_{\text{ddw}}$ is suitable and leads to
\begin{equation}
    \label{eq:2D_axisymmetric_instability_dust_order_half_growth_rate}
    (\gammaI + i \omegaI) \, \tRd = -i \, \langle F_{\text{ddw}}, k_x \, F_{\text{dzf}} \rangle \, \tUgy ,
\end{equation}
where ${ \langle f, g \rangle = \int_{0}^{T_{v}} \frac{\rd s}{T_{v}} \, f^{*} (s) g(s) }$ is the natural Hermitian product over the space of $T_{v}$-periodic functions. This equation captures the forward action described in \S\ref{ssub:2D_instability_why}, by which zonal flows concentrate dust and amplify dust density waves.

\subsection{Order \texorpdfstring{$\dtg$}{} in the gas equations}
\label{sub:Non_modal_RDI_theory_order_1_gas}

\newcommand{\voby}{\overline{v}_{0, y}} 
\newcommand{\hugIIy}{\hat{u}_{g, 1, y}} 

At order $\dtg$, the gas equations become
\begin{align}
    & i \bk \bcdot \hbugII = \,\, 0 , \nonumber \\
    & \begin{multlined}[t]
        (\gammaI + i \omegaI) \, \hbugI + \rd_{t} \hbugII + \bA \, \hbugII = \\
        - i \bk \, \hhII - 2 \bOmega \, \hbugII + \frac{\bvob}{\tau} \hrdO . \hspace{-2.5cm}
    \end{multlined} \nonumber
\end{align}
The only non-trivial equation is the azimuthal momentum equation,
\begin{equation}
    \nonumber 
    (\gammaI + i \omegaI) \, \tUgy \, F_{\text{zf}} + \rd_{t} \hugIIy  + A_{2, 2} \, \hugIIy = \frac{\voby}{\tau} \tRd \, F_{\text{ddw}} ,
\end{equation}
where $A_{2, 2}$ is a strain term from matrix $\bA$. 

To eliminate $\hugIIy$, we reuse the `orthogonal-to-the-image' trick on the operator ${ \rd_{t} + A_{2, 2} \, \bI }$. A suitable function is \smash{${ F_{\text{zf}}^{\dagger} : t \mapsto [ \cos{(\varphi)}^{2} + \alpha^{2} \sin{(\varphi)}^{2} ]^{-1/2} }$}, leading to
\begin{equation}
    \label{eq:2D_axisymmetric_instability_gas_order_1_growth_rate}
    (\gammaI + i \omegaI) \, \tUgy = \frac{1}{\tau} \langle F_{\text{zf}}^{\dagger}, \voby \, F_{\text{ddw}} \rangle \, \tRd .
\end{equation}
This equation captures the backward reaction from \S\ref{ssub:2D_instability_why}, through which dust density waves amplify zonal flows.

\subsection{Bringing everything together}
\label{sub:Non_modal_RDI_theory_combine}

Multiplying Eq.~\eqref{eq:2D_axisymmetric_instability_dust_order_half_growth_rate} and Eq.~\eqref{eq:2D_axisymmetric_instability_gas_order_1_growth_rate} gives
\begin{equation}
    \label{eq:growth_rate_squared_2D}
    (\gammaI + i \omegaI)^{2} = - \frac{i}{\tau} \, \langle F_{\text{ddw}}, k_x \, F_{\text{dzf}} \rangle \, \langle F_{\text{zf}}^{\dagger}, \voby \, F_{\text{ddw}} \rangle .
\end{equation}
This is an important result. Unless the right-hand side is real and negative, this equation indicates that the mode of wavenumber $K_x$ grows exponentially with time. Furthermore, it provides a semi-analytical prediction for the growth rate that is easy to evaluate numerically (the two integrals on the right-hand side are not particularly stiff).

One can show the right-hand side is real. To see this, separate the scalar products' integrals in the middle, 
\begin{align}
    \int_{0}^{T_v} \frac{\rd t}{T_v} f (t) &= \int_{0}^{T_v / 2} \frac{\rd t}{T_v} f (t) + \int_{T_v / 2}^{T_v} \frac{\rd t}{T_v} f (t) , \nonumber \\
    &= \int_{0}^{T_v / 2} \frac{\rd t}{T_v} \bigg[ f(t) + f(T_v - t) \bigg] . \nonumber
\end{align}
Then, consider the symmetry of the different terms in Eq.~\eqref{eq:growth_rate_squared_2D} under the transformation ${ t \rightarrow T_v - t }$: $k_{x}$, $F_{\text{zf}}^{\dagger}$ and $F_{\text{dzf}}$ are conserved, $\voby$ changes sign, and $F_{\text{ddw}}$ is conjugated. Therefore, ${ \langle F_{\text{ddw}}, k_x \, F_{\text{dzf}} \rangle }$ is real, ${ \langle F_{\text{zf}}^{\dagger}, \voby \, F_{\text{ddw}} \rangle }$ is imaginary, and the right-hand side of Eq.~\eqref{eq:growth_rate_squared_2D} is real. 

It is harder to predict the sign, but it turns out to be positive, at least when ${ n = 1 }$ and ${ 4 \leq \alpha \leq 6 }$.

\bsp	
\label{lastpage}
\end{document}